\documentclass{emulateapj}
\slugcomment{Submitted to Astronomical Journal}
\usepackage{amsmath}
\usepackage{lscape}
\usepackage{rotating}
\usepackage{graphicx}
\usepackage{subfig}
\newcommand{\kms}{km s$^{-1}$}
\newcommand{\etal}{et al.}

\shorttitle{ALFALFA Catalog for the Leo Region}
\shortauthors{Stierwalt \etal}

\begin{document}
\title{The Arecibo Legacy Fast ALFA Survey: IX. \\
The Leo Region HI Catalog, Group Membership and the HI Mass Function for the Leo I Group}
\author{Sabrina Stierwalt, Martha P. Haynes\altaffilmark{1}, Riccardo Giovanelli\altaffilmark{1}}\affil{Center for Radiophysics and Space Research, Space Sciences Building, Cornell University, Ithaca, NY 14853.}
\author{Brian R. Kent\altaffilmark{2}}\affil{Jansky Fellow of the National Radio Astronomy Observatory, 520 Edgemont Road, Charlottesville, VA 22903.}
\author{Ann M. Martin}\affil{Center for Radiophysics and Space Research, Space Sciences Building, Cornell University, Ithaca, NY 14853.}
\author{Am\'{e}lie Saintonge}\affil{Institute for Theoretical Physics, University of Zurich, Winterhurerstrasse 190, CH-8057 Zurich, Switzerland.}
\author{Igor D. Karachentsev}\affil{Special Astrophysical Observatory of the Russian AS, Nizhnij Arkhyz 369167, Russia.}
\author{Valentina E. Karachentseva}\affil{Astronomical Observatory of the Kiev Taras Shevchenko National University, Observatorna 3, 04053, Kiev, Ukraine.}
\altaffiltext{1}{National Astronomy and Ionosphere Center, Cornell University, Space Sciences Building, Ithaca, NY 14853. The National Astronomy and Ionosphere Center is operated by Cornell University under a cooperative agreement with the National Science Foundation.}
\altaffiltext{2}{The NRAO is a facility of the National Science Foundation operated under cooperative agreement by Associated Universities, Inc.}


\begin{abstract}
We present the catalog of HI sources extracted from the ongoing Arecibo Legacy Fast ALFA (ALFALFA) extragalactic HI line survey, found within the sky region bounded by $9^{h}36^{m} < \alpha < 11^{h}36^{m}$ and $+08^{\circ} < \delta < +12^{\circ}$. The HI catalog presented here for this 118-deg$^2$ region is combined with ones derived from surrounding regions also covered by the ALFALFA survey to examine the large scale structure in the complex Leo region. Because of the combination of wide sky coverage and superior sensitivity, spatial and spectral resolution, the ALFALFA HI catalog of the Leo region improves significantly on the numbers of low HI mass sources as compared with those found in previous HI surveys. The HI mass function of the Leo I group presented here is dominated by low-mass objects: 45 of the 65 Leo I members have $M_{HI} < 10^{8} M_{\odot}$, yielding tight constraints on the low-mass slope of the Leo I HI mass function. The best-fit slope is $\alpha \simeq -1.41 + 0.2 - 0.1$. A direct comparison between the ALFALFA HI line detections and an optical search of the Leo I region proves the advantage of the ALFALFA strategy in finding low mass, gas-rich dwarfs. These results suggest the existence of a significant population of low surface brightness, gas-rich, yet still very low HI mass galaxies, and may reflect the same type of morphological segregation as is seen in the Local Group. While the low mass end slope of the Leo I HI mass function is steeper than that determined for luminosity functions of the group, the slope still falls short of the values predicted by simulations of structure formation in the $\Lambda$CDM paradigm.
\end{abstract}


\section{Introduction}

In the age of large scale surveys, astronomy has evolved from small number statistics to samplings of large volumes yielding increasingly large datasets. In addition to the characteristics of the stellar population and nuclear activity gleaned from surveys like the Sloan Digital Sky Survey (SDSS, \cite{sdss}), the Two Micron All Sky Survey (2MASS, \cite{2mass}), and the Galaxy Evolution Explorer All Sky Survey (GALEX AIS, \cite{GALEXAIS}), a key ingredient to understanding the formation and evolution of galaxies is their neutral gas content, a tracer of star formation potential and past mergers. Among the first generation HI surveys were the Arecibo HI Strip Survey (AHISS, \cite{AHISS}), the Arecibo Dual Beam Survey (ADBS, \cite{adbs}), and the HI Parkes All-Sky Survey (HIPASS, \cite{hipassother,hipass}). AHISS made 66 HI detections in 65 deg$^2$, and ADBS yielded 265 extragalactic detections in $\sim$430 deg$^2$ of northern sky. HIPASS found 4315 and 1002 objects in the southern and northern catalogs respectively which together covered 71$\%$ of the sky. By combining the superior sensitivity of the Arecibo telescope and its new multi-beam L-band feed array (ALFA), the Arecibo Legacy Fast ALFA Survey (ALFALFA; \cite{PaperI,PaperII}) improves on previous efforts both in its higher angular resolution, in sensitivity and in spectral bandwidth and resolution. The ALFALFA survey aims to obtain a thorough census of neutral hydrogen in the local universe and will eventually cover 7000 deg$^{2}$ of sky over a redshift range of $-$2000 to 18,000 \kms. 

The large areal coverage of the ALFALFA survey and its ability to detect objects of extremely narrow HI line width will permit the determination of the low-mass slope of the HI mass function (HIMF) and its relationship to galaxy environment. ALFALFA is already dramatically increasing the number of known dwarfs in the local universe, including $\sim$300 objects with HI masses below $10^{8}M_{\odot}$ found so far in the catalog extracted from only 20\% of the planned survey. Born out of small mass-density fluctuations, these dwarfs would originally exist as low-mass dark matter halos under the $\Lambda$CDM framework, and, according to hierarchical clustering models, may be analogs of the building blocks of more massive galaxies seen at the current epoch. However, despite strong agreement between simulations of structure formation and large-scale observations like the cosmic microwave background \citep{spergel} and galaxy clustering (e.g. \cite{percival}), the number and distribution of observed dwarf satellites do not match those predicted for the low-mass halos \citep{miss2, misssats}. 

Studies of the luminosity function of galaxies in a range of galaxy environments \citep{lf, cosmos} have consistently found shallower faint end slopes $\alpha$ than the value of $-1.8$ predicted by $\Lambda$CDM \citep{ps74, blumen}. Similar examinations of the low-mass end of the HIMF will determine whether or not these objects instead exist as a population of low-mass, gaseous haloes that either lack stars entirely or are very low surface brightness. We present in this paper the first HIMF dominated by low-mass objects and derived from the ALFALFA dataset: the HIMF for the Leo I group of galaxies.

To probe adequately the low mass slope of the HIMF, the ALFALFA blind HI survey targets the inner regions of the Local Supercluster for the lowest mass galaxies detectable outside of the Local Group. In addition to the area in and around the Virgo cluster \citep{RGcat, BKcat}, early coverage of the ALFALFA survey crosses the Leo region, a complex collection of structures crowded in velocity space and in a small area of sky. In the foreground is Leo I, the nearest group to contain giant ellipticals, lenticulars, and spirals \citep{orig1}. At the Leo I distance of roughly 11 Mpc, ALFALFA can detect objects down to a mass of $\sim 5 \times 10^6$ M$_{\odot}$ for an HI line width of 25 \kms at a signal-to-noise level of 6. Relative to other nearby groups, Leo I is poor in terms of overall luminosity and number of L$^*$ galaxies \citep{lf,fs2}  and is characterized by a low crossing time and velocity dispersion ($\sim 175$ \kms; see Section \ref{veldispsec}). However, the total luminosity of Leo I is still higher than that of the Local Group \citep{pandv} where the local density enhancement is not thought to be large enough to support large and luminous early-type galaxies like the E/S0 galaxies found in Leo I. Intermediate density locations like Leo I where the intragroup medium is typically not dense enough for ram pressure stripping to be significant, but where interactions with other group members clearly occur, are key to understanding the effects of environment on galaxy evolution.

Evidence of interactions among Leo I group members is most obvious in the previously identified extended HI features known as the Leo Ring and in the region of the Leo Triplet. An intergalactic ring of neutral hydrogen roughly 225 kpc in diameter surrounds M105 and NGC~3384 and contains $1.67 \times 10^{9} M_{\odot}$ of HI gas \citep{M96grp}. A spur connects the gas cloud with the warped disk of the M96 spiral galaxy, suggesting the gas may have been swept out after an interaction involving NGC~3384 and M96 \citep{otherringref}. However, the central bars and dust rings of the three brightest nearby galaxies (NGC~3384, M96 \& M105) which all have similar spatial orientations to the Ring \citep{ringref2} and the Ring's Keplerian rotation \citep{M96multi} together suggest the gas could instead be primordial and left over from the formation of the galaxy group. 

Another tidal encounter in Leo I is thought to have occurred between the large spirals NGC~3628 and M66 (NGC~3627) to create one of the largest known tidal tails extending $\sim$100 kpc off of NGC~3628 \citep{tripmodel, maptrip, tripobs}. In M66, an asymmetric HI disk and a recent (less than 1 Gyr ago) burst of  star formation coinciding with the time of the suspected encounter both suggest that an interaction between NGC~3628 and M66 led to NGC~3628's extensive gas loss \citep{tripmodel2, tripother}. Despite being morphologically similar to M66, the third member of the Triplet, NGC~3623, has had a quiescent star formation history in the more recent past and appears to have escaped any direct collisions \citep{tripother}. 

We present in this paper the fifth catalog installment of the ALFALFA survey covering a portion of the Leo region defined here as  $9^{h}36^{m} < \alpha < 11^{h}36^{m}$ and $+08^{\circ} < \delta < +12^{\circ}$. In this 118 deg$^{2}$ of sky, ALFALFA has produced 549 good quality detections. In this same region of the sky, ADBS detected a total of 45 objects, while the northern extension to HIPASS (NHICAT: \cite{hipass}) found only 23 sources. We also take advantage of the availability of survey data in the surrounding regions and present ALFALFA statistics for all detections within $9^{h}36^{m} < \alpha < 11^{h}36^{m}$ and $+04^{\circ} < \delta < +16^{\circ}$. Full catalogs for this additional coverage will be part of future ALFALFA data releases (Haynes \etal ~${\textit{in prep}}$; Giovanelli \etal ~${\textit{in prep}}$). To place the ALFALFA catalog in the context of optical surveys, we have compared the catalog of galaxies derived from the ALFALFA observations with that based on the optical identification of galaxies in the Leo region presented in \cite{KKLeo}. To allow us to compare better the nature of HI selection in Leo, we have also obtained longer integration time, higher sensitivity single-pixel HI observations of each of the optically-selected potential Leo members.

In Section \ref{obs}, we briefly describe the ALFALFA observations, followed by the presentation of the fifth ALFALFA catalog installment. A discussion of the statistics  derived from the currently available ALFALFA catalog in the Leo region is found in Section \ref{overallalf}. We describe the group structure within the Leo region in Section \ref{grouping} and then offer a comparison of an optically- and an HI- selected survey in Section \ref{optselect}. In Section \ref{himfsec}, we present the HI mass function for the group. Finally, we conclude with a discussion of the broader impact of the ALFALFA results in the Leo region.\\

\section{ALFALFA Observations and Data Analysis \label{obs}}
Detailed descriptions of the 2-pass, fixed-azimuth, drift mode strategy exploited by the ALFALFA survey are given in previous papers \citep{PaperI,PaperII,ASb,BKcat,AnnM09}. With the backend correlator set to a bandwidth of 100 MHz spanned by 4096 channels, the resulting spectral resolution is 24.4 kHz ($\sim$5.3 \kms~at a redshift of 0 before the Hanning smoothing that is applied to all of the data presented here). The observations required to construct the dataset presented here were acquired over the months of February and March in 2005, 2006, and 2007.

Once all of the necessary observations are completed, data are gridded into cubes of 2.4$^{\circ}$ by 2.4$^{\circ}$ covering the survey's full spectral bandwith, corresponding to $-$2000 \kms~to +18000 \kms,  with 1$^{\prime}$ sampling. All HI detections determined to be definitely, likely, or possibly real (see Section \ref{catalog} for how detections are ranked) are immediately cross-referenced with SDSS, DSS2, NED, and the Arecibo General Catalog (AGC; a private database of extragalactic objects maintained by M.P.H. and R.G.). The median pointing accuracy, defined here as the difference between the HI centroid and its corresponding optical counterpart, is 25$^{\prime\prime}$ for the lowest signal-to-noise sources ($S/N < 6.5$). With even smaller pointing errors for higher signal-to-noise detections, the corresponding optical galaxy for each HI detection can be identified with a very low margin of error. See Giovanelli \etal ~(${\textit{in prep}}$) for a complete explanation of the gridding and data reduction process.

\subsection{A New ALFALFA Catalog of the Region $09^h36^m < \alpha < 11^h36^m, +08^{\circ} < \delta < +12^{\circ}$ \label{catalog}}
We present in Table \ref{paramstable} the ALFALFA catalog covering $09^{h}36^{m} < \alpha < 11^{h}36^{m}$ and $+08^{\circ} < \delta < +12^{\circ}$. Similar to catalogs presented in earlier ALFALFA data releases, the content for the different columns is as follows: 

\begin{itemize}
\item Col. 1: an entry number for this catalog

\item Col. 2: the source number in the AGC

\item Col. 3: centroid position (J2000) of the HI source after correction for systematic telescope pointing errors (see \cite{BKcat} for a description of how pointing errors vary with declination for the Arecibo telescope). The accuracy of HI positions depends on source strength.

\item Col. 4: centroid position (J2000) of the optical galaxy found to provide the most reasonable optical counterpart to the HI detection. Assignments of optical identifications are made via the ${\it{Skyview}}$ website and are based on spatial proximity, morphology, color, and redshift. Accuracy of centroids is estimated to be $\leq 25 \arcsec$. For cases with lacking or ambiguous optical counterparts, comments are provided as alerted by an asterisk in Col. 14. 

\item Col. 5: heliocentric velocity of the HI source in km s$^{-1}$, cz$_{\odot}$, measured as the midpoint between the channels at which the flux density drops to 50$\%$. The error on cz$_{\odot}$ can be estimated as half the error on the width, as tabulated in Col. 7.

\item Col. 6: velocity width of the source line profile measured at the 50$\%$ level. Corrections for broadening but not turbulent motions, disk inclination, or cosmological effects are applied. In parentheses we show the estimated error on the velocity width, estimated by the sum in quadrature of two components: a statistical error, principally dependent on the S/N ratio of the feature measured, and a systematic error associated with the observer's subjective guess at the quality of the chosen spectral extent of the feature. In the majority of cases, the statistical error is significantly larger than the systematic error; thus the latter is ignored.

\item Col. 7: integrated line flux of the source, ${\it{F_{c}}}$, in Jy km s$^{-1}$. This is measured on the integrated spectrum, obtained by spatially integrating the source image over a solid angle of at least 7$\arcmin\times$7$\arcmin$ and dividing by the sum of the survey beam values over the same set of image pixels (see Shostak \& Allen 1980). 

\item Col. 8: signal-to-noise ratio S/N of the detection, as estimated by
\begin{equation}
S/N = (\frac{1000F_{c}}{W50})\frac{\omega_{smo}^{1/2}}{\sigma_{rms}}
\end{equation}
\noindent where ${\it{F_{c}}}$ is the integrated flux density, as listed in Col. 7, the ratio of $1000F_{c}/W50$ is the mean flux across the feature in mJy, $\omega_{smo}$, the smoothing width expressed as the number of spectral resolution bins of 10 km s$^{-1}$ bridging half of the signal width, is either $W50/(2 \times 10)$ for $W50 < 400$ km s$^{-1}$ or $400/(2 \times 10) = 20$ for $W50 \geq 400$ km s$^{-1}$, and $\sigma_{rms}$ is the r.m.s. noise figure across the spectrum measured in mJy at 10 km s$^{-1}$ resolution, as tabulated in Col. 9.

\item Col. 9: noise figure of the spatially integrated spectral profile, $\sigma_{rms}$, in mJy. The noise figure is the r.m.s. as measured over the signal- and rfi-free portions of the spectrum, after Hanning smoothing to a spectral resolution of 10 \kms.

\item Col. 10: adopted distance in Mpc, $D_{Mpc}$. For objects with $cz_{cmb} >$ 6000, the distance is simply $cz_{cmb}/H_{o}$, where $cz_{cmb}$ is the recessional velocity measured in the Cosmic Microwave Background reference frame and H$_{o}$ is the Hubble constant, for which we use a value of 70 km s$^{-1}$Mpc$^{-1}$. For objects of lower $cz_{cmb}$, we use the multiattractor, peculiar velocity model for the local Universe presented in \cite{KLMthesis}. Objects which are thought to be parts of clusters or groups (for group membership assignments \cite{springob07}) are assigned the $cz_{cmb}$ of the cluster or group. A detailed analysis of group and membership of Leo objects is presented in Section \ref{grouping}.

\item Col. 13: logarithm in base 10 of the HI mass, in solar units. That parameter is obtained by using the expression
$M_{HI} = 2.356 \times 10^{5} D^{2}_{Mpc} F_{c}$.

\item Col. 14: object code, defined as follows:

Code 1 refers to sources of S/N and general qualities that make it a reliable detection: an approximate S/N threshold of 6.5, a good match between the two independent polarizations, and a spatial extent consistent with the characteristics of the telescope beam. Thus, some candidate detections with $S/N \>$ 6.5 have been excluded on grounds of polarization mismatch, spectral vicinity to RFI features or peculiar spatial properties. Likewise, some features of $S/N <$ 6.5 are included as reliable detections if the source's optical characteristics clearly resemble typical galaxies found at the redshift of the HI feature. We estimate that detection candidates with $S/N <$ 6.5 in Table \label{params} will be confirmed in follow-up observations in better than 95$\%$ of cases \citep{ASb}.\\

Code 2 refers to sources of low S/N ($<$ 6.5), which would ordinarily not be considered reliable detections by the criteria set for code 1. However, those HI candidate sources are matched with optical counterparts with known optical redshifts which, within their respective errors, coincide with those measured in the HI line. We refer to these sources as ``priors''.\\

Code 9 refers to objects assumed to be high velocity clouds (HVCs) based on their low heliocentric velocities ($<$200 \kms) and their lack of an optical counterpart; no estimate for their distance is made.\\

Notes flag. An asterisk in this column indicates that a comment is included for this source in the text below.

Only the first few entries of Table 1 are listed in the printed version of this paper. The full content of Table 1 is accessible through the electronic version of the paper and will be made available also through our public digital archive site.\footnotemark \footnotetext{http://arecibo.tc.cornell.edu/hiarchive/alfalfa/} \\

\end{itemize}

\section{Current ALFALFA Dataset for the Leo Region \label{overallalf}}

\begin{figure*}[htp!]
\begin{center}
\includegraphics[width=5in, viewport=15 5 450 580,clip]{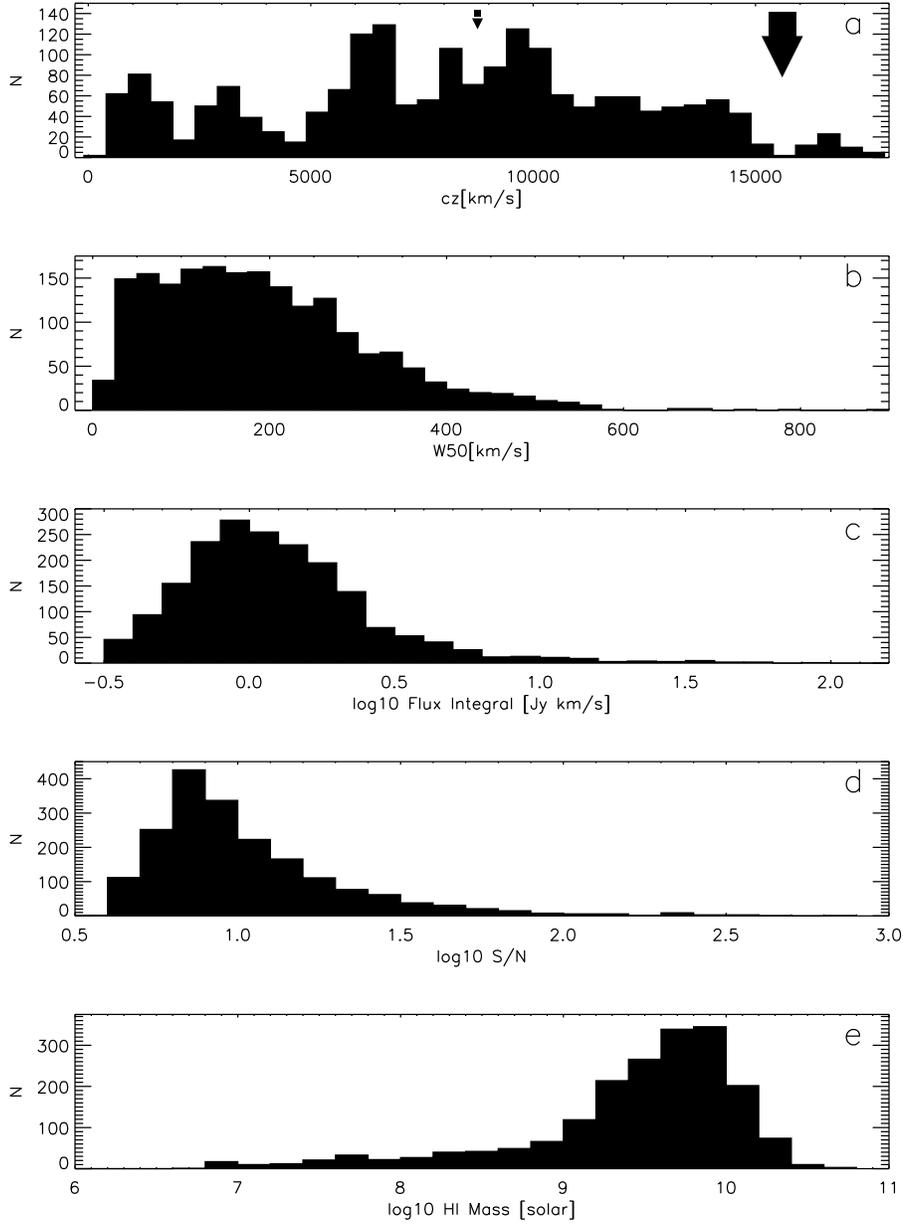}
\caption{Distributions of properties of the sources from the region ($9^h36^m < \alpha < 11^h36^m$ and $+04^{\circ} < \delta < +16^{\circ}$. (a) shows the redshift distribution in
\kms ~with arrows indicating the most significant interruptions due to radio frequency interference (arrow size reflects rfi strength), (b) shows
the velocity width 
distribution
in \kms, (c)
shows the integrated flux distribution in Jy \kms, (d)
shows the S/N distribution, and (e) shows the HI mass distribution in solar mass units.
\label{histplot}}
\end{center}
\end{figure*}

In addition to the catalog presented in Table \ref{paramstable}, the current ALFALFA dataset extends both north and south from $+04^{\circ}$ to $+16^{\circ}$ over the same range of right ascension. To place the catalog data in the context of the surrounding large-scale structure, we use the entire available ALFALFA dataset in the ``Leo region'', defined here as $09^{h}36^{m} < \alpha < 11^{h}36^{m}$ and $+04^{\circ} < \delta < +16^{\circ}$ for the remainder of this paper. The limits in right ascension safely span known Leo I members, while avoiding the Virgo cluster at higher RA, and the limits in declination reflect the currently available survey dataset. Figure \ref{histplot} shows the distributions of sources according to redshift, velocity width, integrated flux, signal-to-noise ratio, and HI mass for this 354 deg$^{2}$ and over the entire survey bandwidth ($-$2000 to 18,000 \kms).
 
The contribution of the Leo group to large scale structure is evident in the spike between cz $\sim 500$ \kms~and cz $\sim 2000$ \kms~in Figure \ref{histplot}a. Also contributing to the structure is the noticeable paucity of sources at a redshift of $\sim$ 2200 \kms~just behind Leo II. The peak in the distribution at velocities just above 3000 \kms~represents the Cancer-Leo Cloud which contains the NGC~3367 group \citep{tully87}. Further out are three Abell clusters (A1016, A999, and A1142) each with about 35 members and all at nearly the same redshift ($cz_{cluster} = $~9600 \kms, 9500 \kms, and 10500 \kms~respectively) that contribute to the peak near 10,000 \kms. Two artificial dips in the histogram result from RFI due to the San Juan FAA radar transmitter at 1350 MHz and its harmonic at 1380 MHz as noted by \cite{RGcat}. The locations and relative strengths of these interferences are represented by downward arrows in Figure \ref{histplot}a. Other RFI contributions are negligible when averaged over the whole dataset.

The significant contribution of the ALFALFA survey to the number of known dwarf galaxies in the nearby universe is revealed in the remaining histograms. Because they are of low mass, dwarf galaxies are expected to have small W50. Most previous HI surveys have been limited by their poorer spectral resolution to detection of significantly larger line widths. For example, no objects with W50 $<$ 30 \kms~were found in all of southern HIPASS ($\sim$21,000 deg$^{2}$; \cite{hipassother}) while, as shown in Figure \ref{histplot}b, 55 low width sources are included in the current ALFALFA catalog in the Leo region alone (354 deg$^2$). Roughly half of these low-W50 HI detections have no associated optical galaxy and are thought to be emission from either the Leo Ring or the extended HI in the Leo Triplet region (see Section \ref{nooc}). Of the remaining 30 sources, only ten have signal-to-noise ratios of greater than 10 suggesting that, while ALFALFA clearly has the ability to detect objects of very narrow line widths, dwarf galaxies with W50$<$ 30 \kms~are most likely rarer than dwarfs of higher W50. 

New detections of nearby dwarfs are also expected to push the lower limits of HI flux and mass. As shown in Figure \ref{histplot}c, nearly half of the objects from the current ALFALFA catalog of the Leo region have integrated fluxes of less than 1.0 Jy \kms, the completeness limit for broad signals in ALFALFA. (The limit is even lower for narrow signals at $\sim $0.25 Jy \kms.) In more than 60 times the areal sky coverage, the southern HIPASS catalog contains only one source below 1.0 Jy \kms ~\citep{hipassother}. The Leo sample also reaches down to $M_{HI} = 10^{6.77} M_{\odot}$ as shown in Figure \ref{histplot}e. Of the 1953 good quality detections in the sample, 118 have $M_{HI} < 10^{8}M_{\odot}$ (roughly 6$\%$), and 45 of these low-mass galaxies were determined to be members of Leo I (see Section \ref{grouping}). The percentage of low-mass objects in the Leo region is comparable to the 8$\%$ found in the much denser Virgo region \citep{RGcat, BKcat}. Although Virgo at 16 Mpc is larger and more densely populated, Leo is slightly closer to the Local Group at $\sim$11 Mpc away and thus allows for the detection of even lower mass objects down to $5\times 10^6 M_{\odot}$.

\begin{figure*}[htp!]
\plotone{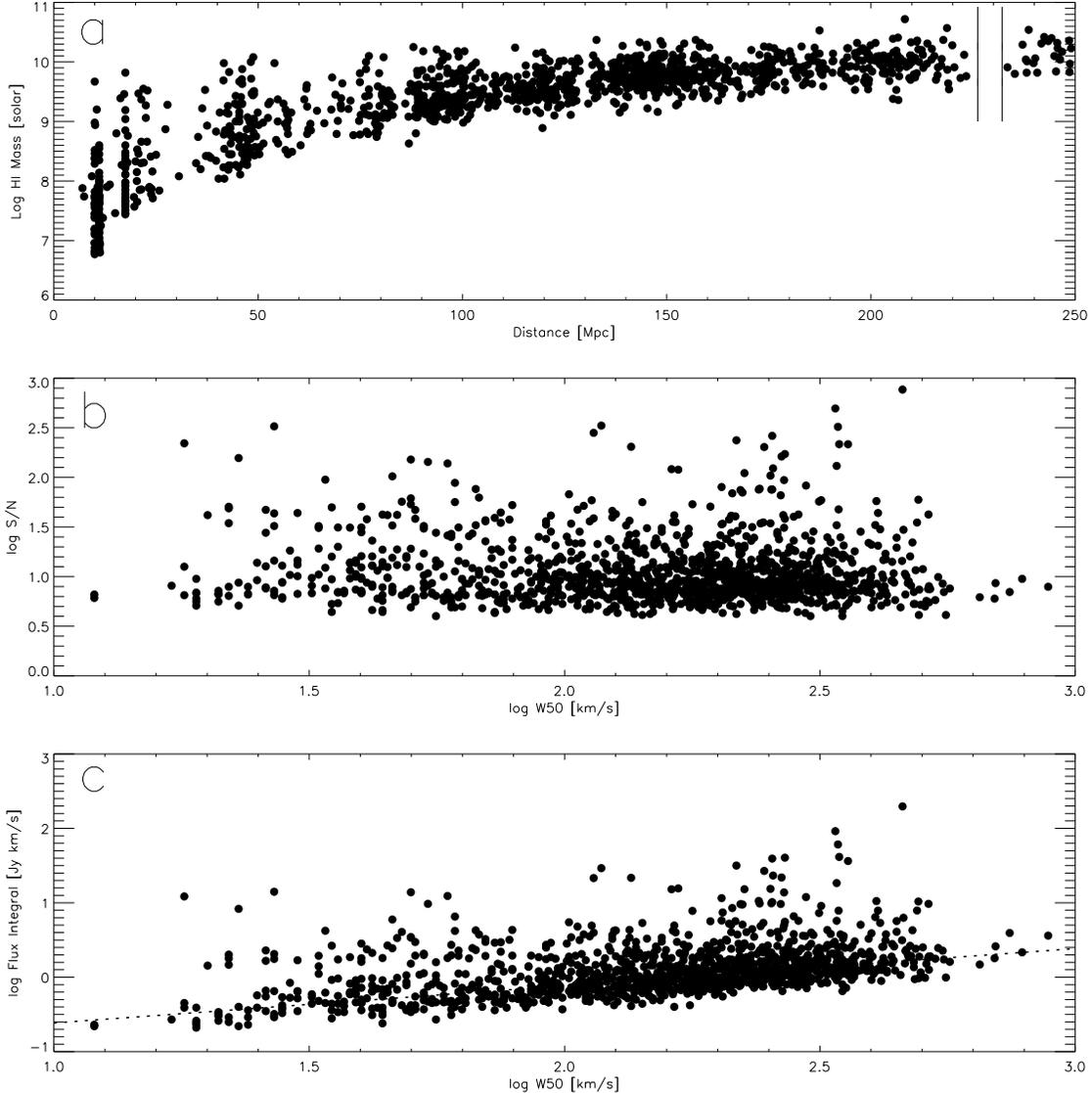}
\caption{Statistical properties of the sources from the region ($9^{h}36^{m} < \alpha < 11^{h}36^{m}$
and $+04^{\circ} < \delta < +16^{\circ}$).
The upper panel shows the logarithm of HI mass versus distance. The gap in detections between 220 and 230 Mpc is due
to RFI while the gap in sources at 30 Mpc reflects the corresponding large-scale structure. 
The middle panel shows the logarithm of S/N versus the logarithm of velocity width, W50. The lower
envelope is constant over the entire W50 range. The lower panel shows the logarithm of integrated flux versus the logarithm of velocity width. Here the lower envelope is dependent on width and the dashed line indicates a S/N level of 6.5. 
\label{dotplot}}
\end{figure*}

Figure \ref{dotplot} shows the relation of HI mass to distance and of S/N and integrated flux to velocity width for the sample. The stacking of objects at 11 and 17 Mpc in Figure \ref{dotplot}a comes from placing nearby objects at a variety of recessional velocities at the distances adopted for Leo I and II (see Section \ref{grouping} for an explanation of how the distances and group members were chosen). The lack of sources seen in Figure \ref{histplot}a around 2200  \kms~(which roughly translates to 30 Mpc) is still present. This paucity is also seen in the distribution of the $\sim$3800 optical redshifts found in the AGC within the same right ascension, declination, and velocity bounds, suggesting the gap accurately reflects the large scale structure in the region. RFI from the San Juan FAA radar transmitter causes the gap in detections near 230 Mpc shown both by the vertical lines in the upper panel of Figure \ref{dotplot} and by the dip at 16,000 km s$^{-1}$ in the redshift histogram of Figure \ref{histplot}a.

The middle and lower panels of Figure \ref{dotplot} show that while the distribution of S/N appears to be unbiased toward larger velocity widths (i.e. the lower envelope is constant over the entire width range), the integrated flux values do depend on width. The dashed line in the lower panel indicates a constant S/N level of 6.5. This expected trend \citep{PaperII} was also noted in previous ALFALFA catalog releases \citep{RGcat, AScat, BKcat, AnnM09}. Only seven objects have W50 $>$ 600 \kms, and only one of the seven (UGC 6066) has a $S/N > 10$ (UGC 6066, an edge-on galaxy with $cz = 11,807$ \kms~\& W50 $= 667$ \kms). The low number of high-W50 detections in this dataset is partially a reflection of the small area of sky being considered; nineteen additional high-W50 sources are found in the currently available ALFALFA spring-sky catalog, six of which have $S/N > 10$. Despite more high-W50 sources found in the ALFALFA-Leo catalog than in the southern HIPASS one (only 8 objects with W50 $>$ 600 \kms; \cite{hipassother}), the number of galaxies falls off quickly for large widths, W50 $>$ 600 \kms~for both samples. Although an intrinsic rarity of sources of high velocity width and low flux most likely plays a role in this paucity, such objects are also difficult to detect both by eye and with matched-filtering algorithms for fixed rms, resulting in a known survey bias against large widths. (See Section \ref{himfsec} for completeness corrections applied to compensate for this bias.)

\subsection{Leo Features without Optical Counterparts \label{nooc}}

More than 100 detections listed in the ALFALFA catalogs for the Leo region do not have clearly associated optical counterparts. Half of these can be linked to either the Leo Ring, the extended HI in the Leo Triplet, or tidal remnants surrounding NGC~3389. In this section we present detailed maps of these three systems. Another 43 objects are nearby high velocity clouds (denoted by a code `9'). Of the remaining 15 detections with no optical counterparts, 14 have marginal signal to noise and require further followup. The most promising candidate for an independent system without a detectable stellar component is AGC 215416 at a cz of 3371 \kms. With a signal-to-noise ratio of 17, the putative HI detection is placed at a distance of 50 Mpc by the \cite{KLMthesis} flow model and thus at an HI mass of $10^{8.75}M_{\odot}$ but has no visible emission in either SDSS or DSS2 blue images. One very likely possibility is that AGC 215416 is an OH megamaser at z$\sim$0.19. Deeper optical and HI observations are needed to determine the nature of this object.

\subsection{ALFALFA Survey Map of the Leo Ring}\label{ringsec}

Since its serendipitous discovery \citep{ringdet}, the Leo Ring has been studied at 21-cm (Arecibo: \cite{ringdet}; VLA: \cite{VLAring}), molecular (CO \& OH: \cite{M96multi}), infrared (IRAS: \cite{M96multi}), optical (V and K bands: \cite{ringVK}; R band: \cite{ringR}; B \& V bands: \cite{M96multi}; H$\alpha$: \cite{ringHa}), and X-ray frequencies \citep{M96multi}. Other than a tentative, 4$\sigma$ H$\alpha$ detection by \cite{ringHa}, until recently only neutral hydrogen searches in the intergalactic cloud have been successful. However, new GALEX observations \citep{thilker} have revealed ultraviolet emission possibly associated with star formation within the Ring's neutral gas. 

\cite{M96grp} found $2.06 \times 10^9 M_{\odot}$ of HI in the Ring with an integrated flux of $S_{int} = 70.9$ Jy \kms~for the distance of 11.1 Mpc adopted here. We identify 26 separate clumps which constitute the Ring, yielding a total HI mass of $1.80 \times 10^9 M_{\odot}$ (24$\%$ of the $M_{HI}$ for the entire M96 group), an integrated flux of $S_{int} = 62.14$ Jy \kms, and a mean velocity of 852 \kms. The ALFALFA flux budget may not account for some of the low surface brightness components of the Ring, so the 12\% flux mismatch is not a source of serious concern. The most massive contribution to the Ring, containing the spur connecting to M96, as well as the additional structures to the north, east, and west found in the \cite{M96multi} map are all recovered by the ALFALFA dataset, and no new significant structures are found despite the much larger sky coverage of the ALFALFA map. Both the \cite{M96multi} map and the ALFALFA one show a velocity gradient with lower velocities found for the more scattered clumps to the northeast and higher velocities belonging to the larger portions in the southwest.

\begin{figure}[htp]
\begin{center}
\includegraphics[width=3.3in]{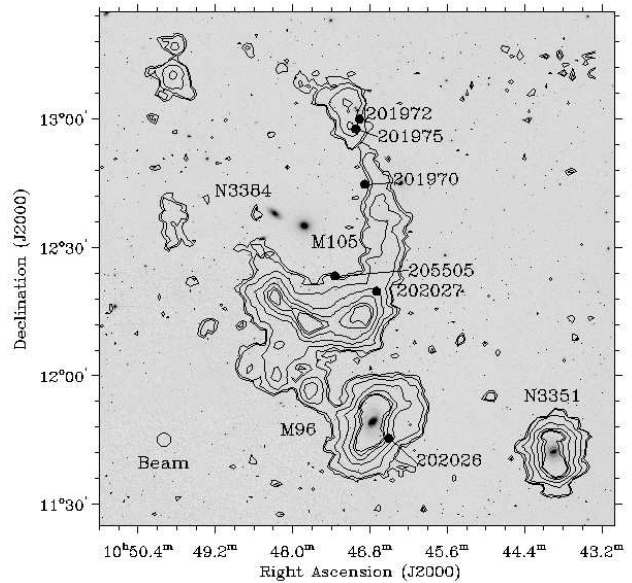}
\caption{The map of the Leo Ring extracted from the ALFALFA dataset over the velocity range 708 \kms~to 1046 \kms, overlaid on a mosaic of SDSS r-band images. HI contours are drawn at 4.0, 5.0, 9.0, 18, 32, 44, and 50 mJy per beam (units are left in mJy per beam as some of the emission is resolved). The open circle represents the ALFA HPBW of $\sim$4$^{\prime}$. Dwarf galaxies noted in optical surveys of the region that lie within the extent of the Ring are shown as filled circles and labeled with their AGC number. Optical redshifts are not known for AGC 202026, AGC 201975, and AGC 201972, so the HI detections at these locations cannot be differentiated from Ring emission and may not be associated with the optical galaxies. For AGC 202027 and AGC 201970, optical redshifts are known that match the measured 21-cm line velocities. These two dwarf galaxies may have formed from overdensities in the Ring. AGC 205505 was not identified by the signal extraction algorithm, and an optical redshift of 1146$\pm$50 \kms~places the optical galaxy at this position just above the range of velocities covered by the Ring. The association of AGC 205505 with the Ring is thus more tenuous. The largest optical galaxies (N3351, N3368=M96, N3384, and N3379=M105) are labeled.\label{ringcont}}
\end{center}
\end{figure}

The ALFALFA survey map of the Ring covering 708 \kms ~to 1046 \kms ~is shown in Figure \ref{ringcont} overlaid on a mosaic of SDSS r-band images. The largest optical galaxies in the M96 group are NGC~3384 and NGC~3379=M105 at the center of the Ring, NGC~3368=M96 which is connected to the Ring by a spur of HI, and NGC~3351 to the west. Six optically-identified galaxies found superimposed on the Ring and thus possibly associated with the HI are noted by filled circles and labeled with their AGC numbers. Three of the optical detections, AGC 201972 (called KK94 in \cite{KK98}), and AGC 202026 and AGC 201975 from \cite{KKLeo}, have unknown optical redshifts. Thus the HI detected at those locations by ALFALFA cannot be definitively linked to the optical galaxy, but instead may be associated with the Ring. Higher resolution synthesis HI observations and optical redshifts are needed to ascertain the nature of these detections. 

In the cases of AGC 202027 and AGC 201970, optical redshifts are reported that match those measured in the HI observations; SDSS gives a redshift of 1013 \kms~for AGC 202027, and \cite{KKLeo} cite a redshift of 617 \kms~for AGC 201970. For AGC 205505, an independent HI detection was not identified by the signal extraction algorithm \citep{ASb}, and SDSS finds a redshift for the associated optical galaxy of 1146$\pm$50 \kms, which is slightly higher than the range of HI velocities covered by emission from the Ring. Thus AGC 205505 is considered an M96 group member but the association of the optical galaxy with the Ring is more tenuous. These three dwarf galaxies may have formed out of overdensities in the Ring. A study of their metallicities would determine whether they formed with the other, more massive Leo I group members which may mean the Ring is primordial as suggested by \citep{M96grp} or whether they are high metallicity tidal dwarf systems which may have resulted from the tidal encounter that produced the Ring.

Two additional optical detections noted by \cite{KKLeo} as potential Leo I members that overlap with the Ring, AGC 200592 and AGC 201963, were found to be background sources with optical redshifts of 16,775 \kms~and 53,213 \kms~respectively, and are not marked. For details on the comparison of the optically-selected \cite{KKLeo} sample with the ALFALFA catalog throughout the rest of the M96 group, see Section \ref{optselect}).

\subsection{ALFALFA Survey Map of the Leo Triplet}
Less than 1.5 Mpc away from the Leo Ring is the trio of large spiral galaxies NGC~3623, NGC~3627, and NGC~3628 collectively known as the Leo Triplet. \cite{optplume} noted a faint optical plume extending from NGC~3628 directed eastward away from the other two galaxies. Subsequent 21-cm observations found an associated neutral hydrogen plume roughly 100 kpc in length, an HI bridge connecting NGC~3628 to NGC~3623, and a distortion in the HI disk of NGC~3627 \citep{tripmodel, maptrip}.

Despite the appearance of a bridge of gas connecting NGC~3623 to the perturbed NGC~3628, simulations of a collision that is prograde for NGC~3627 and retrograde for NGC~3628 but not involving NGC~3623 match best with the observations of the plume \citep{toomre, tripmodel, maptrip}. NGC~3623 also has a relatively quiescent star formation history \citep{tripother}, further suggesting that NGC~3623 was not involved in perturbing NGC~3628. The dust-to-gas ratio in the plume, determined by observations of the plume's infrared component, is also consistent with the tidal model for its formation \citep{plumemod}. No CO has yet been detected in the plume \citep{coplume}.

\begin{figure}[htp]
\begin{center}
\includegraphics[width=3.3in]{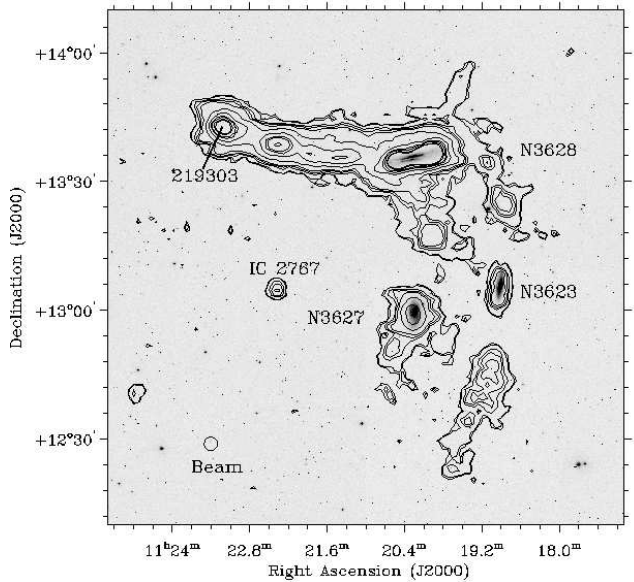}
\caption{The map of the Leo Triplet derived from the ALFALFA dataset over the velocity range 631 \kms~to 1150 \kms, overlaid on a mosaic of SDSS r-band images. HI contours are drawn at 4.5, 5.0, 8.0, 10, 13, 26, 52, 78, 91, 117, and 130 mJy per beam (units are left in mJy per beam as some of the emission is resolved). The open circle represents the ALFA HPBW of $\sim$4$^{\prime}$. All four ALFALFA detections in the field associated with optical galaxies (N3623=M65, N3627=M66, N3628, and IC 2767) are labeled. The location of the optical galaxy AGC 219303 which is possibly associated with the plume but has no optical redshift is indicated.\label{tripcont}}
\end{center}
\end{figure}

The ALFALFA map of the Triplet is shown in Figure \ref{tripcont} covering 631 \kms~to 1150 \kms~and overlaid on a mosaic of SDSS r-band images. The main features of the \cite{maptrip} map of the region are all recovered: the large plume extending eastward of NGC~3628, a clump of  HI between NGC~3628 to NGC~3623 (M65), and the cloud extending to the southwest of NGC~3627. The ALFALFA dataset detects $1.0 \times 10^9 M_{\odot}$ in the plume which contributes 14$\%$ of the entire HI mass for the M66 group. The same amount of gas mass is reported in \cite{maptrip} for the group distance of 10.0 Mpc adopted here. The earlier map finds a roughly constant velocity field over 50 kpc of the plume's length at $\sim$900 \kms. Although the ALFALFA dataset shows emission throughout the plume at that velocity, the ALFALFA map also reveals gas along the entire plume with a larger range of 860 \kms ~$< v <$ 920 \kms~as well as gas at low relative velocities ($\sim$840 \kms) found only at the far end of the plume. 

Additional HI located outside the area covered by previous HI observations is found by ALFALFA to extend south of NGC~3627 and northward from NGC~3628. The ALFALFA detection registers $2.3 \times 10^8 M_{\odot}$ in the southern HI clump which is 3$\%$ of the entire HI mass for the M66 group. We identify eight separate clouds within the southwestern clump and they are presented in Table \ref{M66table}. Lower velocities ($\sim$650 \kms) dominate the southernmost part of the clump, and velocity increases as the gas is traced upward until matching the HI velocity of NGC~3627 at 750 \kms. This newly-detected HI and its spatial and spectral proximity to NGC~3627 furthers the hypothesis that NGC~3627 was responsible for the collision that led to the plume. The ALFALFA data do not reveal any disturbance in the velocity field of NGC~3623 which may suggest it was not involved in the past encounter. The HI clump that appears to connect NGC~3623 with NGC~3628 is actually well separated in velocity from NGC~3623. Thus it is not a bridge between the two galaxies, but instead an extension of NGC~3628.

In Figure \ref{tripcont}, we note the location of an optical galaxy seen in POSS-II and SDSS (AGC 219303). The irregular galaxy is very low surface brightness with a B band apparent magnitude of 17.5 and has a morphology consistent with other Leo I dwarfs. An optical redshift is needed to determine whether AGC 219303 is associated with the HI in the plume. 

\subsection{ALFALFA Survey Map of the NGC~3389 System}\label{n3389sec}
Located just behind the Leo Ring at a c$z$ of 1301 \kms, the large SA galaxy, NGC~3389 shows a large central peak in its HI profile, possibly a sign of a prior tidal encounter \citep{M96grp}. However, \cite{orig0} claimed NGC~3389 was not part of the M96 group, and thus unlikely to be interacting with the Ring. After limited mapping of a few points surrounding NGC~3389, \cite{M96grp} found a nearby dwarf, CGCG 066-029 (AGC 200603), to be the most likely cause of NGC~3389's centrally peaked profile based on the dwarf's unusual morphology and a slight extension of its HI toward NGC~3389. \cite{hoff1987} believed CGCG 066-029 to be part of a binary pair with AGC 200604 with only a 30 \kms ~difference in velocity. Neither \citep{M96grp} nor \cite{hoff1987} reported a connection with the dwarf 20 arcminutes to the south, CGCG 066-025 (AGC 200598).

\begin{figure}[htp]
\begin{center}
\includegraphics[width=3.3in]{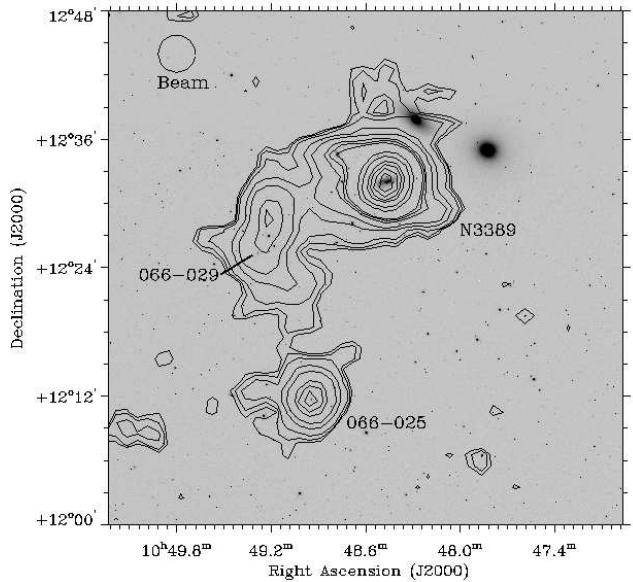}
\caption{Map of the region around NGC~3389 derived from the ALFALFA dataset over the velocity range 1123 \kms~to 1487 \kms, overlaid on an SDSS r-band image. HI contours are drawn at 0.75, 1.0, 1.45, 2.9, 4.4, 7.3, 8.7, 10, 20, 26, 35, 40, 45, and 50 mJy per beam (units are left in mJy per beam as some of the emission is resolved). The open circle represents the ALFA HPBW of $\sim$4$^{\prime}$. The three optical galaxies in the system (N3389, CGCG 066-025, and CGCG 066-029) are labeled. The bright S0 galaxy, NGC~3384, and the bright elliptical galaxy, NGC~3379, seen to the northwest of NGC~3389 are foreground galaxies located in the center of the Leo Ring (see Figure \ref{ringcont}).\label{n3389cont}}
\end{center}
\end{figure}

The ALFALFA map of the area surrounding NGC~3389 is shown in Figure \ref{n3389cont} covering 1123 \kms~to 1487 \kms~and overlaid on a mosaic of SDSS r-band images. A clear connection is seen in position and velocity space between NGC~3389 and CGCG 066-029. A $2^{\prime}$ displacement is revealed in the direction of NGC~3389 between the centroid of the HI and the stellar component of CGCG 066-029 which furthers the idea that the two galaxies are interacting. However, AGC 200604, originally noted as the binary partner of CGCG 066-029 \citep{hoff1987}, does not appear to be part of the system at all, and the optical galaxy lines up with an ALFALFA HI detection at 6941 \kms~instead. A bridge connecting CGCG 066-025 to the rest of the system is clearly detected suggesting that despite its smooth looking contours, this second dwarf may also be involved in the interaction. This bridge contains $6.3 \times 10^7 M_{\odot}$ adopting the secondary distance to NGC~3389 of 21.4 Mpc (see Table \ref{LIItable}). As noted by \cite{M96grp} the NGC~3389 system may be similar to the tidal encounter between the Magellanic Clouds and the Milky Way where the dwarfs are distorted by the close encounter with a large spiral neighbor. The only minimally disturbed gas distribution and morphology of CGCG 066-025 may suggest the galaxy became involved in the encounter on more recent timescales.

\section{Structure and Dynamics of the Leo Group}\label{grouping}

The Local Group of galaxies is part of the Local Sheet, a plane-like structure with a spread of only $\sim$1.5 Mpc about the supergalactic plane SGZ=0 \citep{T08}. The nearest adjacent structure is Leo I (also called the Leo Spur),  a complex grouping of galaxies over a narrow velocity range (roughly 500 \kms ~$<$ v $<$ 1200 \kms). Since the earliest references to the Leo group of galaxies \citep{orig1, orig2}, studies have noted the existence of substructure within the Leo I group, most commonly the M96 group including the Leo Ring and the M66 group including the Leo Triplet \citep{orig3, hg}. Some authors have further separated Leo I into even more distinct groups \citep{materne, tully87}, but others suggest that velocity crowding due to the proximity of the Virgo cluster leads to the appearance of more group structure than may actually exist \citep{M96grp}. Adding further confusion to the Leo I group structure is the more disperse Leo Cloud in the background. \citep{tully87}. Like Leo I, the Leo Cloud is most likely an assemblage of several smaller groups.

As different authors tend to use the same nomenclature to refer to different groups, the definitions we use are presented in Table \ref{definitiontable}. The centroid position, mean velocity (with error), velocity dispersion (with error), number of members, harmonic mean radius, and assumed distance are listed for each group or subgroup. For the M96 and M66 groups, the centroid positions are M96 and M66 themselves. The center of the background Leo II group is placed approximately along a line of sight between M96 and M66. The mean velocities, velocity dispersions, assumed distances, and number of members are determined in Sections \ref{groupmemI}, \ref{distances}, and \ref{groupmemII}. The harmonic mean radius is calculated using the distance to each group, the number of members, and the angular separation of those members.

\begin{figure}[h!]
\begin{center}
\vspace{0pt}
\hspace{0pt}
\includegraphics[width=3.5in]{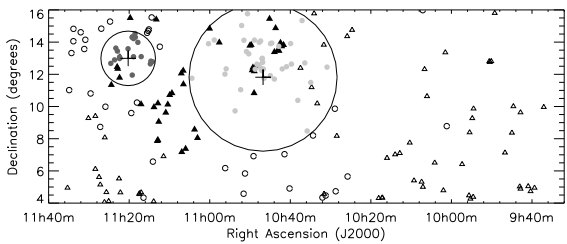}
\caption{Membership in the Leo Region. Filled symbols represent probable group members: dark gray circles for the M66 group (which includes the Leo Triplet), light gray circles for the M96 group (which includes the Leo Ring), and triangles for Leo II. Open symbols represent objects with less clearly defined group memberships: open circles are likely Leo I members and open triangles are probably part of the Leo Cloud. M96 and M66 are marked as large crosses for reference. Large circles surround the M96 and M66 groups at radii of 0.8 Mpc and 0.3 Mpc respectively.
\label{vis}}
\end{center}
\end{figure}

\begin{deluxetable*}{llcrrrcrl}
\tablewidth{0pt}
\tabletypesize{\scriptsize} 
\tablecaption{Hierarchical Structures in the Leo Region\label{definitiontable}}
\tablehead{
\colhead{Cloud~~~~~} & \colhead{Group} & \colhead{Centroid} & \colhead{$\bar{v}$ ($\epsilon_{v}$)} & \colhead{$\sigma$
($\epsilon_{\sigma}$)} &
\colhead{~~~$R_{H}$} & 
\colhead{Dist} & \colhead{Members} & \colhead{Notes} \\
 & {Name} & {(J2000)} & {\kms} & {\kms} & {Mpc} & {Mpc} & &
}
\startdata
Leo I & & - & 906 (26) & 203 (15) & $>$0.27& - & 96~~~~ & also called Leo Spur\\
 & M96 & 10 46 45.7 +11 49 12 & 881 (16) & 174 (19) & 0.13 & 11.1 & 41~~~~ & part of Leo I\\
 & M66 & 11 20 15.0 +12 59 30 & 812 (12) & 156 (26) & 0.06 & 10.5 & 19~~~~ & part of Leo I\\
Leo Cloud & & - & $\sim$1500~~ & $\sim$200~~ & $>$0.50 & - & $>$100~~~~ & behind Leo I\\
 & Leo II & $\sim$11 03 +12 24 & 1423 (24) & 181 (20) & 0.22 & 17.5 & 41~~~~ & part of Leo Cloud\\
\hline
\enddata
\end{deluxetable*}

\subsection{Group Membership in Leo I}\label{groupmemI}

Nearest neighbor searches and other group finding algorithms for determining group membership are easily confused by the high density of sources and complicated group structure in Leo. Both Leo I and the Leo Cloud are projected on the same small area of sky at very similar redshifts and thus are difficult to separate. Due to Virgo's proximity, the Leo I group's infall velocity to Virgo of $> 300$ \kms~\citep{sakai} can counteract pure Hubble flow significantly and thus confuse redshift measurements. Leo I also has at least two distinct components at nearly the same redshift: the M96 and M66 groups. For M96 and M66 to be part of the same bound structure, the crossing time for such a group would have to be $\tau_{cross} = 2R/\sigma = 1.7\times 10^{10}$ years, given a $\sigma$ of 172 km s$^{-1}$ (as calculated in Section \ref{veldispsec}). Thus Leo I has not had time to virialize and can still be split into two entities to better understand its dynamics. 

For group membership statistics in Leo I, we thus rely on both velocity dispersion calculations and spatial density distributions to determine group assignments. Potential members are pulled from the AGC; that database includes all detections in the current ALFALFA catalog made to date. By experimenting with different velocity cutoffs as a requirement for group membership (i.e. only galaxies within a certain range of velocities can be deemed members), we find the velocity dispersion rises steeply when galaxies with velocities lower than 600 \kms~or higher than 1200 \kms~are included. Thus we require a source velocity of $600$ \kms ~$< cz < 1200$ \kms~for a galaxy to be considered a member of Leo I. Next, by placing M96 and M66 at the centers of separate Leo I subgroups, a radius for each subgroup is found beyond which the number of additional group members levels off. We choose a radius of 1.7$^{\circ}$ for the M66 group  or $\sim$ 0.3 Mpc at the M66 distance of 10.0 Mpc and a radius of 4.3$^{\circ}$ for the M96 group or $\sim$ 0.8 Mpc at the M96 distance of 11.1 Mpc (group distances are determined in Section \ref{distances}). The group radius determined for the M96 group does not change if NGC~3384 (found at the center of the Leo Ring) is made the group center instead of M96.

After this analysis, the M96 group is found to have 39 members (not including the detections that make up the Ring) and the M66 group to have 19 members. Included in the M96 members are all of the optically-identified dwarfs from \cite{KKLeo} that were spectroscopically confirmed as Leo I members (see Section \ref{optselect}). Our membership designations also include all galaxies named as M96 or M66 group members in the Nearby Optical Galaxy catalog of nearby groups (NOG, \cite{NOG}). 

The HI properties for these objects are summarized in Tables \ref{M96table} and \ref{M66table}. In the M96 group, 26 HI detections with no optical counterparts are attributed to the Leo Ring. In the M66 group, 22 objects are listed without optical counterparts: sixteen are attributed to either the plume of HI gas extending from NGC~3628 or the extended clump of HI just south of NGC~3627. The remaining six objects are not connected to any of the Triplet galaxies via HI bridges above the ALFALFA survey detection limit, although association is likely. These six detections are treated separately and labeled `HIonly' in Table \ref{M66table}. Parameters for all objects are taken from the ALFALFA catalog unless otherwise noted, and not all objects have associated HI detections. 

\begin{deluxetable*}{llcccccc}
\tablewidth{0pt}
\tabletypesize{\scriptsize} 
\tablecaption{Leo I - M96 Group Membership \label{M96table}}
\tablehead{
\colhead{AGC} & \colhead{Other} & \colhead{Opt Position\tablenotemark{a}} & \colhead{HI cz$_{\odot}$} & \colhead{W50} & \colhead{$F_c$} &
\colhead{Dist\tablenotemark{b}} & \colhead{$\log M_{HI}$}
     \\
 {\#} & {~Name} & {(J2000)} & {\kms} & {\kms} & {Jy \kms} & {Mpc} & {$M_{\odot}$}
} 
\startdata
205156 &         & 10 30 52.9 +12 26 48 &  ~915 &  ~~21 &   0.32 & 11.1 & 6.91 \\
202248 &         & 10 34 56.1 +11 29 32 & 1177 &  ~~62 &   0.64 & 11.1 & 7.28 \\
202017\tablenotemark{c} & LeG03   & 10 35 48.9 +08 28 49 & 1158 &  ~~70 &   1.93 & 11.1 & 7.75 \\
  5761 & N3299   & 10 36 24.0 +12 42 24 &  ~604 & 112 &   3.54 & 11.1 & 8.00 \\
205165 &         & 10 37 04.8 +15 20 15 &  ~724 &  ~~27 &   0.30 & 11.1 & 6.93 \\
200499 & 065-065 & 10 38 08.0 +10 22 51 & 1175 & 178 &   7.79 & 11.1 & 8.35 \\
202019\tablenotemark{c} & LeG05   & 10 39 43.0 +12 38 04 &  ~780 &  ~~22 &   0.08 & 11.1 & 6.37 \\
200512\tablenotemark{c} & LeG06   & 10 39 55.6 +13 54 34 & 1007 &  ~~21 &   0.28 & 11.1 & 6.91 \\
  5812 & 065-083 & 10 40 56.5 +12 28 18 & 1008 &  ~~56 &   1.59 & 11.1 & 7.65 \\
200532 & 065-086 & 10 42 00.3 +12 20 07 &  ~772 &  ~~36 &   0.96 & 11.1 & 7.46 \\
205268 &         & 10 42 52.4 +13 44 28 & ~~~~~~~1145 (opt) & ~~... & ... & 11.1 & ... \\
  5850 & N3351   & 10 43 57.6 +11 42 12 &  ~777 & 270 &  40.41~~ & 10.0\tablenotemark{*} & 8.98 \\
205445 &         & 10 44 35.3 +13 56 23 &  ~~~~~~~~633 (opt) & ~~... & ... & 11.1 & ... \\
200560\tablenotemark{d} &         & 10 44 54.6 +13 54 29 & 1010 &  ~~29 &   0.61 & 11.1 & 7.25 \\
202024\tablenotemark{c} & LeG13   & 10 44 57.3 +11 55 01 &  ~871 &  ~~24 &   0.22 & 11.1 & 6.81 \\
202026\tablenotemark{c} & FS 15   & 10 46 30.2 +11 45 19 &  ~954 & 126 &   3.24 & 11.1 & 7.97 \\
205287 & Ring    & 10 46 36.0 +12 37 44 &  ~957 &  ~~78 &   3.18 & 11.1 & 7.94 \\
205289 & Ring    & 10 46 36.4 +12 26 02 & 1006 &  ~~48 &   4.06 & 11.1 & 8.06 \\
202027\tablenotemark{c} & FS 17   & 10 46 41.3 +12 19 37 & 1030 &  ~~37 &   1.24 & 11.1 & 7.56 \\
205290 & Ring    & 10 46 42.4 +12 46 56 &  ~915 &  ~~50 &   1.52 & 11.1 & 7.63 \\
  5882 & N3368   & 10 46 45.7 +11 49 11 &  ~893 & 343 &  60.81~ & 10.5\tablenotemark{*} & 9.20 \\
201970\tablenotemark{c} & LeG18   & 10 46 52.2 +12 44 40 &  ~636 &  ~~38 &   0.55 & 11.1 & 7.20 \\
201972 & KK94    & 10 46 57.3 +12 59 53 &  ~834 &  ~~33 &   1.94 & 11.1 & 7.75 \\
201975\tablenotemark{c} & LeG21   & 10 47 00.8 +12 57 34 &  ~843 &  ~~23 &   0.48 & 11.1 & 7.14 \\
205291 & Ring    & 10 47 02.7 +12 13 36 & 1018 &  ~~50 &  13.86~ & 11.1 & 8.60 \\
205292 & Ring    & 10 47 09.1 +13 03 11 &  ~824 &  ~~27 &   1.76 & 11.1 & 7.71 \\
205293 & Ring    & 10 47 19.1 +13 09 30 &  ~806 &  ~~51 &   0.37 & 11.1 & 7.02 \\
205505 &         & 10 47 20.1 +12 23 15 & ~~~~~~~1146 (opt) & ~~... & ... & 11.1 & ... \\
  5889 & N3377A  & 10 47 22.4 +14 04 14 &  ~573 &  ~~46 &   5.96 &  9.3\tablenotemark{*} & 8.08 \\
205294 & Ring    & 10 47 39.1 +11 55 52 &  ~971 &  ~~27 &   2.05 & 11.1 & 7.79 \\
  5899 & N3377   & 10 47 42.3 +13 59 08 &  ~~~~~~~~689 (opt) & ~~... & ... & 11.2\tablenotemark{*} & ... \\
205295 & Ring    & 10 47 47.8 +12 13 07 &  ~978 &  ~~59 &  12.36~ & 11.1 & 8.55 \\
  5902 & N3379   & 10 47 49.6 +12 34 55 &  ~~~~~~~~911 (opt) & ~~... & ... & 11.0\tablenotemark{*} & ... \\
205296 & Ring    & 10 47 49.7 +13 07 47 &  ~787 &  ~~30 &   0.52 & 11.1 & 7.20 \\
205297 & Ring    & 10 48 04.3 +13 11 24 &  ~794 &  ~~21 &   0.26 & 11.1 & 6.93 \\
205301 & Ring    & 10 48 12.2 +12 04 14 &  ~927 &  ~~47 &   3.36 & 11.1 & 8.00 \\
205302 & Ring    & 10 48 13.6 +12 08 38 &  ~917 &  ~~46 &   2.45 & 11.1 & 7.76 \\
205303 & Ring    & 10 48 15.6 +12 18 02 &  ~910 &  ~~54 &   9.69 & 11.1 & 8.40 \\
  5911 & N3384   & 10 48 16.8 +12 37 42 &  ~~~~~~~~728 (opt) & ~~... & ... & 11.6\tablenotemark{*} & ... \\
205304 & Ring    & 10 48 28.1 +12 25 53 &  ~854 &  ~~98 &   1.17 & 11.1 & 7.37 \\
205305 & Ring    & 10 48 30.4 +12 37 43 &  ~648 &  ~~44 &   0.91 & 11.1 & 7.40 \\
205306 & Ring    & 10 48 32.8 +12 30 07 &  ~794 &  ~~75 &   0.70 & 11.1 & 7.23 \\
205307 & Ring    & 10 48 36.0 +12 02 56 &  ~924 &  ~~27 &   0.65 & 11.1 & 7.28 \\
205308 & Ring    & 10 48 42.8 +13 16 05 &  ~785 &  ~~12 &   0.22 & 11.1 & 6.82 \\
200596 & 066-026 & 10 48 53.7 +14 07 27 &  ~~~~~~~~637 (opt) & ~~... & ... & 11.1 & ... \\
205311 & Ring    & 10 49 12.3 +12 11 51 &  ~869 &  ~~18 &   0.39 & 11.1 & 7.01 \\
201963\tablenotemark{e} & Ring    & 10 49 51.3 +13 09 24 &  ~766 &  ~~20 &   1.43 & 11.1 & 7.60 \\
205313 & Ring    & 10 49 51.5 +12 36 49 &  ~774 &  ~~30 &   0.64 & 11.1 & 7.25 \\
205314 & Ring    & 10 49 51.9 +13 17 21 &  ~787 &  ~~18 &   0.45 & 11.1 & 7.12 \\
205315 & Ring    & 10 49 52.4 +12 32 21 &  ~779 &  ~~33 &   0.48 & 11.1 & 7.13 \\
205316 & Ring    & 10 49 56.7 +12 40 22 &  ~776 &  ~~45 &   0.62 & 11.1 & 7.31 \\
205321 & Ring    & 10 50 02.6 +13 06 30 &  ~788 &  ~~19 &   0.25 & 11.1 & 6.74 \\
205322 & Ring    & 10 50 09.2 +13 00 30 &  ~797 &  ~~23 &   0.22 & 11.1 & 6.79 \\
  5944 & 064-033 & 10 50 18.9 +13 16 18 & ~~~~~~~1073 (opt) & ~~... & ... & 11.1\tablenotemark{*} & ... \\
  5948 &         & 10 50 38.2 +15 45 48 & 1121 & 106 &   4.79 & 11.1 & 8.14 \\
  5952 & N3412   & 10 50 53.2 +13 24 42 &  ~~~~~~~~867 (opt) & ~~... & ... & 11.3\tablenotemark{*} & ... \\
205540 &         & 10 51 31.4 +14 06 53 &  ~~~~~~~~832 (opt) & ~~... & ... & 11.1 & ... \\
205544 &         & 10 52 04.8 +15 01 50 &  ~~~~~~~~828 (opt) & ~~... & ... & 11.1 & ... \\
202456 &         & 10 52 19.5 +11 02 36 &  ~~~~~~~~824 (opt) & ~~... & ... & 11.1 & ... \\
  6014 & 066-058 & 10 53 42.7 +09 43 39 & 1133 &  ~~94 &   2.90 & 11.1 & 7.92 \\
202034\tablenotemark{c} &         & 10 55 55.3 +12 20 22 &  ~847 &  ~~22 &   0.10 & 11.1 & 6.46 \\
202035\tablenotemark{c} & D640-13 & 10 56 13.9 +12 00 37 &  ~989 &  ~~30 &   1.67 & 11.1 & 7.69 \\
205278 &         & 10 58 52.2 +14 07 46 &  ~686 &  ~~36 &   0.34 & 11.1 & 7.01 \\
  6082 & N3489   & 11 00 18.6 +13 54 04 &  ~695 & 113 &   0.70 & 12.1\tablenotemark{*} & 7.36 \\
210023 & 066-109 & 11 04 26.3 +11 45 21 &  ~777 &  ~~44 &   1.81 & 11.1 & 7.70 \\
\hline
\enddata
\tablenotetext{a}{Positions indicate the centroid of the optical counterpart unless the object is noted as a Ring detection, in which case the position represents the centroid of the HI.}
\tablenotetext{b}{Objects are assigned a group distance except when a * indicates a known primary distance.}
\tablenotetext{c}{HI parameters come from single pixel results (presented in Section \ref{optselect})}
\tablenotetext{d}{HI parameters come from previously catalogued single-pixel Arecibo observations. See the HI archive for details.}
\tablenotetext{e}{Ring detections; optical redshifts from SDSS place optical sources at these locations as background galaxies; ADBS gives v=754 \kms~for AGC 201963}
\end{deluxetable*}

\begin{deluxetable*}{llcccccc}
\tablewidth{0pt}
\tabletypesize{\scriptsize} 
\tablecaption{Leo I - M66 Group Membership \label{M66table}}
\tablehead{
\colhead{AGC} & \colhead{Other\tablenotemark{a}} & \colhead{Opt Position} & \colhead{HI cz$_{\odot}$} & \colhead{W50} & \colhead{$F_c$} &
\colhead{Dist\tablenotemark{b}} & \colhead{$\log M_{HI}$} 
     \\
{\#} & {~Name} & {(J2000)} & {\kms} & {\kms} & {Jy \kms} & {Mpc} & {$M_{\odot}$} 
} 
\startdata
215387 & HIonly  & 11 14 14.5 +12 46 55 &  ~578 &  ~~75 &   2.20 & 10.0 & 7.73 \\
  6272 & N3593   & 11 14 37.0 +12 49 02 &  ~631 & 254 &   9.84 & 10.0 & 8.36 \\
202256 &         & 11 14 45.0 +12 38 51 &  ~630 &  ~~42 &   0.64 & 10.0 & 7.16 \\
210220 & I2684   & 11 17 01.1 +13 05 55 &  ~588 &  ~~25 &   0.57 & 10.0 & 7.09 \\
215386 & HIonly  & 11 17 50.6 +13 59 06 &  ~871 &  ~~27 &   0.32 & 10.0 & 6.86 \\
215389 & HIonly  & 11 18 28.2 +14 18 13 &  ~917 &  ~~28 &   0.39 & 10.0 & 6.95 \\
215392 & HIonly  & 11 18 33.1 +14 32 02 &  ~909 &  ~~17 &   0.27 & 10.0 & 6.79 \\
215393\tablenotemark{d} & Plume   & 11 18 52.4 +13 24 33 &  ~862 &  ~~40 &   1.80 & 10.0 & 7.59 \\
215397 & HIonly  & 11 18 54.4 +14 13 07 &  ~909 &  ~~22 &   0.29 & 10.0 & 6.73 \\
215396\tablenotemark{c} & SClump  & 11 18 53.6 +12 53 50 &  ~581 &  ~~25 &   0.39 & 10.0 & 6.97 \\
  6328 & N3623   & 11 18 55.7 +13 05 32 &  ~803 & 493 &  10.42 & 10.0 & 8.37 \\
215398\tablenotemark{c} & SClump  & 11 19 05.2 +12 45 28 &  ~753 &  ~~41 &   2.37 & 10.0 & 7.77 \\
215400\tablenotemark{c} & SClump  & 11 19 08.0 +12 39 16 &  ~753 &  ~~26 &   1.66 & 10.0 & 7.62 \\
215401\tablenotemark{d} & Plume   & 11 19 11.8 +13 35 43 &  ~834 &  ~~49 &   1.06 & 10.0 & 7.35 \\
215286 &         & 11 19 12.7 +14 19 40 &  ~998 &  ~~28 &   0.54 & 10.0 & 7.12 \\
215354 &         & 11 19 15.9 +14 17 25 &  ~~~~~~~~728 (opt) & ~~... & ... & 10.0 & ... \\
215402\tablenotemark{d} & Plume   & 11 19 25.7 +13 14 12 &  ~772 &  ~~82 &   1.02 & 10.0 & 7.39 \\
215403\tablenotemark{c} & SClump  & 11 19 30.1 +12 33 57 &  ~716 &  ~~98 &   1.20 & 10.0 & 7.43 \\
215405\tablenotemark{c} & SClump  & 11 19 33.0 +12 31 00 &  ~695 &  ~~76 &   1.19 & 10.0 & 7.50 \\
215406\tablenotemark{d} & Plume   & 11 19 33.5 +13 51 44 &  ~984 &  ~~74 &   1.52 & 10.0 & 7.53 \\
215407\tablenotemark{c} & SClump  & 11 19 37.4 +12 23 44 &  ~655 &  ~~26 &   0.66 & 10.0 & 7.20 \\
215409\tablenotemark{c} & SClump  & 11 19 54.0 +12 52 40 &  ~678 &  ~~47 &   0.68 & 10.0 & 7.17 \\
215410\tablenotemark{d} & Plume   & 11 19 58.6 +13 17 33 &  ~785 &  ~~40 &   2.83 & 10.0 & 7.82 \\
  6346 & N3627   & 11 20 15.0 +12 59 21 &  ~720 & 359 &  36.44~ & 10.0\tablenotemark{*} & 8.92 \\
  6350 & N3628   & 11 20 16.9 +13 35 13 &  ~844 & 459 & 197.24~~~ & 10.0 & 9.66 \\
215411\tablenotemark{c} & SClump  & 11 20 26.8 +12 52 13 &  ~646 &  ~~67 &   2.32 & 10.0 & 7.43 \\
215412\tablenotemark{d} & Plume   & 11 21 47.4 +13 37 17 &  ~908 &  ~~23 &   8.29 & 10.0 & 8.29 \\
215413\tablenotemark{d} & Plume   & 11 22 23.1 +13 38 55 &  ~905 &  ~~18 &  12.18~ & 10.0 & 8.46 \\
211370 & I2767   & 11 22 23.2 +13 04 40 & 1083 &  ~~92 &   1.75 & 10.0 & 7.62 \\
213436 &         & 11 22 24.0 +12 58 46 &  ~~~~~~~~626 (opt) & ~~... & ... & 10.0 & ... \\
  6395 & I2782   & 11 22 55.5 +13 26 26 &  ~~~~~~~~999 (opt) & ~~... & ... & 10.0 & ... \\
215414\tablenotemark{d,e} & Plume   & 11 23 11.1 +13 42 30 &  ~878 &  ~~27 &  14.09~ & 10.0 & 8.52 \\
  6401\tablenotemark{f} &         & 11 23 19.1 +13 37 45 &  ~883 &  ~~49 &   0.94 & 10.0 & 7.35 \\
213440 & I2791   & 11 23 37.6 +12 53 45 &  ~666 &  ~~22 &   0.25 & 10.0 & 6.67 \\
215415 & HIonly  & 11 24 33.9 +12 40 48 & 1004 &  ~~19 &   0.39 & 10.0 & 6.96 \\
\hline
\enddata
\tablenotetext{a}{Positions indicate the centroid of the optical counterpart unless the object is noted as a plume, southern clump, or HI-only detection, in which case the position represents the centroid of the HI.}
\tablenotetext{b}{Objects are assigned a group distance except when a * indicates a known primary distance.}
\tablenotetext{c}{Components of the HI cloud just south of N3627 are attributed to N3627 for HIMF determination.}
\tablenotetext{d}{Components of the HI plume in the Leo Triplet are attributed to N3628 for HIMF determination.}
\tablenotetext{e}{possible association with very low surface brightness galaxy at 112313.5+134254 (AGC 219303) found in POSS-II and SDSS}
\tablenotetext{f}{HI parameters come from previously catalogued single-pixel Arecibo observations. See the HI archive for details.}
\end{deluxetable*}

\begin{deluxetable*}{llcccccc}
\tablewidth{0pt}
\tabletypesize{\scriptsize} 
\tablecaption{Probable Leo I members \label{LeoItable}}
\tablehead{
\colhead{AGC} & \colhead{Other} & \colhead{Opt Position} & \colhead{HI cz$_{\odot}$} & \colhead{W50} & \colhead{$F_c$} &
\colhead{Dist\tablenotemark{a}} & \colhead{$\log M_{HI}$} 
     \\
{\#} & {~Name} & {(J2000)} & {\kms} & {\kms} & {Jy \kms} & {Mpc} & {$M_{\odot}$} 
} 
\startdata
202171 &         & 10 01 09.5 +08 46 56 & ~~~~~~~1167 (opt) & ~~... & ... & 19.0 & ... \\
  5453 & 093-047 & 10 07 07.2 +15 59 01 &  ~839 &  ~~53 &   1.99 & 13.0 & 7.88 \\
203913 & 037-033 & 10 25 46.4 +05 39 13 & 1155 &  ~~99 &   2.77 & 18.8 & 8.35 \\
208394 &         & 10 28 43.8 +04 44 04 & 1181 &  ~~27 &   0.54 & 19.2 & 7.72 \\
202218 &         & 10 28 55.8 +09 51 47 & 1190 &  ~~39 &   0.59 & 19.6 & 7.73 \\
  5708 & 037-061 & 10 31 13.2 +04 28 19 & 1176 & 169 &  30.24~ & 19.1 & 9.41 \\
204139 &         & 10 32 01.3 +04 20 46 & 1147 &  ~~68 &   0.45 & 18.6 & 7.54 \\
202222 &         & 10 34 21.1 +08 11 56 &  ~~~~~~~~854 (opt) & ~~... & ... & 12.4 & ... \\
208399 &         & 10 40 10.7 +04 54 32 &  ~747 &  ~~23 &   1.00 &  ~9.9 & 7.34 \\
205078 &         & 10 41 26.1 +07 02 16 & 1175 &  ~~32 &   0.42 & 19.4 & 7.58 \\
  5923 & 038-022 & 10 49 07.5 +06 55 01 &  ~709 & 142 &   3.11 &  ~9.0 & 7.76 \\
  5962 & N3423   & 10 51 14.4 +05 50 22 & 1008 & 156 &  32.83~ & 11.7\tablenotemark{*} & 9.02 \\
  5974 & 038-032 & 10 51 35.1 +04 34 57 & 1038 & 155 &  16.57~ & 25.1\tablenotemark{*} & 9.39 \\
200688 & 038-054 & 10 56 09.1 +06 10 22 & 1014 & 128 &   0.69 & 16.8 & 7.67 \\
213066 &         & 11 12 23.2 +13 42 49 &  ~~~~~~~~630 (opt) & ~~... & ... &  ~7.6 & ... \\
211261 & I678    & 11 14 06.3 +06 34 37 &  ~~~~~~~~968 (opt) & ~~... & ... & 13.3 & ... \\
215282 &         & 11 14 25.2 +15 32 02 &  ~867 &  ~~27 &   0.29 & 11.3 & 6.91 \\
  6277 & N3596   & 11 15 06.2 +14 47 12 & 1193 & 118 &  29.22~ & 20.7 & 9.47 \\
215281 &         & 11 15 16.2 +14 41 55 & ~~~~~~~1092 (opt) & ... & ... & 19.0 & ... \\
215284 &         & 11 15 32.4 +14 34 38 & 1133 &  ~~23 &   0.40 & 19.7 & 7.54 \\
212132 & 039-094 & 11 16 26.3 +04 20 11 & 1104 & 155 &   2.15 & 18.6 & 8.24 \\
213006 &         & 11 18 03.9 +10 14 40 &  ~~~~~~~~957 (opt) & ~~... & ... & 12.7 & ... \\
202257 &         & 11 19 14.4 +11 57 07 &  ~861 &  ~~51 &   2.97 & 10.7 & 7.90 \\
213074 &         & 11 19 28.1 +09 35 44 &  ~990 &  ~~51 &   1.95 & 13.7 & 7.93 \\
215142 &         & 11 24 44.5 +15 16 32 & 1125 & 123 &   2.27 & 20.0 & 8.29 \\
  6438 & I692    & 11 25 53.5 +09 59 13 & 1156 &  ~~50 &   3.46 & 20.5 & 8.53 \\
215296 &         & 11 26 55.2 +14 50 03 &  ~913 &  ~~44 &   0.57 & 11.5 & 7.23 \\
210340 & I2828   & 11 27 11.0 +08 43 53 & 1046 &  ~~45 &   2.67 & 17.9 & 8.30 \\
213091 &         & 11 29 34.6 +10 48 36 &  ~~~~~~~~743 (opt) & ~~... & ... &  ~8.6 & ... \\
212837 & KKH68   & 11 30 52.9 +14 08 44 &  ~880 &  ~~22 &   1.79 & 10.7 & 7.68 \\
215303 &         & 11 31 08.8 +13 34 14 & 1021 &  ~~32 &   0.54 & 15.0 & 7.43 \\
215304 &         & 11 32 01.9 +14 36 39 & 1124 & 115 &   1.46 & 20.3 & 8.13 \\
215306 &         & 11 33 50.1 +14 49 28 & 1129 &  ~~64 &   0.45 & 20.4 & 7.54 \\
215248 &         & 11 33 50.9 +14 03 15 &  ~928 &  ~~19 &   0.21 & 11.3 & 6.88 \\
210459 & I2934   & 11 34 19.3 +13 19 18 & 1195 &  ~~61 &   4.19 & 21.4 & 8.65 \\
212838 & KKH69   & 11 34 53.4 +11 01 10 &  ~881 &  ~~22 &   1.47 & 10.4 & 7.57 \\
\hline
\enddata
\tablenotetext{a}{Objects are assigned flow model distances except when a * indicates a known secondary distance.}
\end {deluxetable*}

Thirty-six galaxies remain potential Leo I members but their group membership is unclear. These objects fall within the velocity range of $600$ \kms~$< cz <  1200$ \kms~but are outside the group radii determined for the M96 and M66 subgroups. Group distances are not assigned to these objects, and instead distances determined for each individual galaxy by the Masters (2005) flow model are adopted. The parameters for these sources are found in Table \ref{LeoItable} and are taken from the ALFALFA catalog unless otherwise noted.

\subsection{Primary and Secondary Distances in Leo}\label{distances}

\indent Primary distances are key to placing galaxy groups like Leo I into the larger context of the surrounding large-scale structure. The primary distances known for the Leo I group are listed in Table \ref{primarytable}. Heliocentric radial velocities as quoted in the NASA Extragalactic Database\footnote{http://nedwww.ipac.caltech.edu/} and a primary distance with estimated error are given for each galaxy, as well as the method used to obtain the distance and the reference for the measurement. \cite{freedman} used Cepheid variables, \cite{MK98} measured bright stars, and \cite{tonry} and \cite{rekolasbf} both studied surface brightness fluctuations. 

\indent Nine of the ten primary distance measurements quoted in Table \ref{primarytable} belong to members of the M96 group. The last entry in Table \ref{primarytable} is a Cepheid distance to M66 (UGC 6346/NGC~3627) and is the only primary distance measurement to a member of the M66 group. These primary distances are also reported in \cite{T08}, as well as additional secondary distances determined via the Tully-Fisher relation. Potential Leo members with secondary distances are shown in Table \ref{secondarytable} as calculated from the distance moduli reported in \cite{T08}. 

For our analysis of Leo I, including both the M96 and M66 groups, we adopt the same distances chosen by \cite{T08}: 11.1 Mpc to the M96 group and 10.0 Mpc to the M66 group. These distances represent a weighted average of the known primary distances in the M96 group (as well as the only primary distance in the M66 group) and agree well with the several distance moduli quoted for Leo I in \cite{fs}. Group distances are assigned to all members as 11.1 Mpc for the M96 group and 10.0 Mpc for the M66 group unless a primary distance to the object has been measured. Although we favor group distances over secondary distance measurements, the Tully-Fisher distances are used as a check on the adopted group distances. 

\begin{deluxetable*}{clcccll}[t!]
\tablewidth{0pt}
\tabletypesize{\scriptsize} 
\tablecaption{Primary Distances in Leo I \label{primarytable}}
\tablehead{
\colhead{AGC} & \colhead{Other} & \colhead{Opt Position} & \colhead{~v$_{\odot}$} &
\colhead{Dist $~(\epsilon_{dist})$} & \colhead{Method} & \colhead{Reference~~~~~~~~~~~~~~~~~~~~~~~~~~~~~~~~~~~}\\
{\#} & {~Name} & {(J2000)} & {~\kms} & {Mpc} & & 
}
\startdata
5850 & N3351 & 10 43 57.6 +11 42 12 & ~~778 & 10.00 (0.92) & ceph & Freedman \etal ~2001\\
5882 & N3368 & 10 46 45.7 +11 49 11 & ~~897 & 10.47 (0.96) & ceph & Freedman \etal ~2001\\ 
5889 & N3377A & 10 47 22.3 +14 04 13 & ~~573 & ~9.30 (1.93) & stars & Makarova and Karachentsev 1998\\ 
5899 & N3377 & 10 47 42.3 +13 59 08 & ~~665 & 11.22 (0.47) & sbf & Tonry \etal ~2001\\ 
5902 & N3379 & 10 47 49.6 +12 34 55 & ~~911 & 10.57 (0.54) & sbf & Tonry \etal ~2001\\
5911 & N3384 & 10 48 16.8 +12 37 42 & ~~704 & 11.36 (0.75) & sbf & Tonry \etal ~2001\\ 
5944 & 064-033 & 10 50 18.9 +13 16 18 & 1073 & 11.10 (0.90) & sbf & Rekola \etal ~2005\\
5952 & N3412 & 10 50 53.2 +13 24 42 & ~~841 & 11.32 (0.73) & sbf & Tonry \etal ~2001\\
6082 & N3489 & 11 00 18.6 +13 54 04 & ~~677 & 12.08 (0.83) & sbf & Tonry \etal ~2001\\
6346 & N3627 & 11 20 15.0 +12 59 21 & ~~727 & 10.05 (0.69) & ceph & Freedman \etal ~2001\\
\hline
\enddata
\end{deluxetable*}

\begin{deluxetable*}{clccc}
\tablewidth{0pt}
\tabletypesize{\scriptsize} 
\tablecaption{Secondary Distances\tablenotemark{*}~~in Leo I \& II\label{secondarytable}}
\tablehead{
\colhead{AGC} & \colhead{Other} & \colhead{Opt Position} & \colhead{v$_{\odot}$} &
\colhead{Dist $~(\epsilon_{dist})$}\\
{\#} & {Name} & {(J2000)} & {\kms} & {Mpc}
}
\startdata
5271 & N3020 & 09 50 06.6 +12 48 49 & 1440 & 21.88 (4.03)\\
5303 & N3041 & 09 53 07.2 +16 40 40 & 1408 & 23.77 (4.38)\\
5325 & N3049 & 09 54 49.7 +09 16 18 & 1455 & 15.28 ($\sim$3.5)\\
5914 & N3389 & 10 48 28.0 +12 31 59 & 1308 & 21.38 (3.55)\\
6167 & N3526 & 11 06 56.8 +07 10 27 & 1420 & 19.77 (3.64)\\
6209 & N3547 & 11 09 55.9 +10 43 13 & 1579 & 18.11 (3.34)\\
6328 & N3623 & 11 18 55.7 +13 05 32 & ~~807 & 11.97 (1.93)\\
6387 & IC2763 & 11 22 18.4 +13 03 54 & 1569 & 16.60 ($\sim$3.8)\\
6420 & N3666 & 11 24 26.1 +11 20 32 & 1060 & 16.29 (2.63)\\
6498 & N3705 & 11 30 07.4 +09 16 36 & 1018 & 17.22 (2.78)\\
6594 & U6594 & 11 37 38.3 +16 33 18 & 1038 & 21.28 (3.92)\\
6644 & N3810 & 11 40 58.8 +11 28 17 & ~~993 & 15.35 (2.54)\\
\hline
\enddata
\tablenotetext{*}{All secondary distances are calculated from the distance moduli reported in Tully \etal~2008 from the Tully-Fisher method.}
\end{deluxetable*}

\subsection{Group Membership in the Leo Cloud}\label{groupmemII}
The extent and substructure of the slightly more distant Leo Cloud are less clearly defined than those of Leo I. In their catalog of Tully-Fisher distances, \cite{T08} consider potential Leo Cloud members  spanning over 50 degrees of right ascension and 45 degrees of declination. Due to the limited declination range of the current ALFALFA catalog and the loose association of the galaxies in the expansive Leo Cloud, our search does not cover the entire structure. In fact, 57 of the 72 objects ($\sim 80\%$) marked as Leo Cloud members in \cite{T08} are outside of the current ALFALFA catalog declination range. 

Potential members of the Leo Cloud within the ALFALFA survey limits are found by their velocity. The group velocity dispersion as a function of chosen velocity cut-off begins to rise more steeply after a cz of 2000 \kms, so any object listed in the AGC within $9^{h}36^{m} < \alpha < 11^{h}36^{m}$ and $+04^{\circ} < \delta < +16^{\circ}$ and having $1200$ \kms~$< cz < 2000$ \kms~is considered a potential Leo Cloud member. The HI parameters for these 103 objects are summarized in Table \ref{LIItable} where HI parameters come from the ALFALFA catalog unless otherwise noted. 

\indent Although these potential Leo Cloud members are likely to be associated within large scale structure, sources over such a large expanse of sky cannot all be confidently placed at the same group distance. Instead we focus on the substructure within the Leo Cloud directly behind Leo I on the sky which we define as the Leo II group. Choosing an approximate group center located along the line of sight midway between the M96 and M66 groups, the number of additional Leo II members plateaus for group radii larger than 1.1 Mpc. 41 sources are found to be potential Leo II members within a group radius of 1.1 Mpc. The Leo II group includes the NGC 3389 system (see section \ref{n3389sec}).

No primary distances are known for Leo II, so we use as a reference the nine primary distance estimates placing members of the M96 group firmly at 11.1 Mpc. We adopt a distance of $D =$ 11.1 Mpc $\times (\bar{v}_{II}/\bar{v}_{M96}) = 17.5$~Mpc, where $\bar{v}_{II}$ and $\bar{v}_{M96}$ are the mean velocities of suspected Leo II and M96 group members respectively. This value agrees well with the eight potential Leo II members for which \cite{T08} measured Tully-Fisher distances (see Table \ref{secondarytable}). Distances to all other Leo Cloud sources are estimated via the flow model unless a secondary distance is known.

\subsection{Velocity Dispersion for Leo I}\label{veldispsec}

Previous optical and redshift surveys of Leo I have been plagued by interloping background galaxies which have led to large estimates for the group's velocity dispersion. Using grouping algorithms to search for overdensities in the CfA Redshift Survey, \cite{gh} found the M96 group (their Group $\#$68) and the M66 group (their Group $\#$78) to have 23 and 9 members respectively. If all of their redshift measurements are weighted equally, they determine a velocity dispersion of 258 \kms~for the Leo I group as a whole. However, when compared to the brightest members of each group (M96 and M66), the velocity distribution of group members is skewed towards higher velocities. Six of the 23 M96 members have recessional velocities less than M96 while only one of the 11 M66 members has a velocity below that of M66. In an examination of optical plates from the Las Campanas Observatory, \cite{fs} selected 52 members for the M96 group based primarily on morphology as they have redshifts for only 11 of their assigned M96 members. When equally weighted, the 11 Leo group redshifts result in a velocity dispersion of 256 \kms~which is similar to that of \cite{gh} and thus potentially too high an estimate as well.

By limiting the M96 and M66 groups in right ascension and declination, as well as paying close attention to sudden leaps in the groups' velocity dispersions with additional members, we minimize the problem of background contamination. A look at the 39 M96 members and 19 M66 members (not including any Ring, Clump or Plume detections) we find that roughly half of the member galaxies have velocities less than those of the brightest members (M96 and M66). From the group memberships given in Tables \ref{M96table} and \ref{M66table}, and using ALFALFA-derived heliocentric velocities, we calculate a velocity dispersion for Leo I of 172 \kms~which excludes the sources whose membership status in Leo I is unclear (see Table \ref{LeoItable}). Including these additional 36 sources, however, only raises the velocity dispersion to 181 \kms. If our cutoff velocity for group membership ($v_{cut} = 1200$ \kms) is applied to the groups defined in \cite{gh} and in \cite{fs}, their velocity dispersions are reduced to 136 \kms~and 98 \kms~respectively.

The distribution of Leo I and Leo Cloud members in right ascension and declination is shown in Figure \ref{vis}. Filled dark and light gray circles represent members of the M66 and M96 groups respectively. Open circles denote the probable Leo I members found in Table \ref{LeoItable}. M96 and M66 themselves are marked with large crosses, and Leo II members are plotted as small, open triangles. The large circles indicate the group radii of 0.3 Mpc and 0.8 Mpc determined for the M66 and M96 subgroups respectively.

\section{Comparison with the KK04 Optically-Selected Catalog} \label{optselect}
The Leo region was included in the optically-selected Catalog of Neighboring Galaxies (\cite{KKLeo}; hereafter KK04). To complement the blind HI data from ALFALFA, higher sensitivity, single-pixel HI observations were made with the Arecibo telescope and the higher gain L-band (L-band wide: LBW) receiver and multi-bit autocorrelation spectrometer of 35 dwarf galaxies optically identified by KK04 as potential M96 group members. KK04 selected the objects based on visual scrutiny of POS-II/ESO plates; the pointed observations made with the LBW receiver supplied additional redshift information, and, in some cases, confirmation of Leo I group status. The dwarfs range in B-band magnitude from 19.2 to 17.0 and are found in the 160 deg$^2$ bounded by $10^{h}30^{m} < \alpha < 11^{h}05^{m}$ and $+08^{\circ} < \delta < +16^{\circ}$. This entire region has been included in the ALFALFA survey, and so a direct comparison of optical- and HI-selected galaxies can be made.

Due to instrumental errors, three spectrometer configurations were used resulting in different spectral resolutions. The sources AGCs 202016, 202027, 202035, 200512, 201963, 202028, 201991, 202030, 202032, 202018, 202022, 201990, 202031, and 202038 were observed with a spectral resolution of 12.2 kHz (roughly 2.7 \kms~and 4.5 \kms~at a redshift of 0 before and after Hanning smoothing). The remaining 21 sources were observed with a spectral resolution of 24.4 kHz (roughly 5.3 \kms~and 8.8 \kms~at cz of 0 before and after hanning smoothing).

All 35 targets were initially observed in total power position-switched mode for 120 seconds on source followed by 120 seconds off source, and nine were immediately detected. The remaining 26 targets were observed for longer periods, with final total integration times ranging from 240 seconds to 1440 seconds on source. Nine additional sources were detected with these longer integration times. All spectra were Hanning smoothed, bandpass calibrated, and when available, both polarizations were averaged. For the sources that were detected, the HI spectra were fit with polynomials, and the central velocity, $cz_{\odot}$, the full width at half the signal's maximum height $W_{50}$, and the total integrated flux under the profile $S_{tot}$ were measured. The rms noise level was calculated for all nondetections to allow the estimation of upper limits to their HI mass and HI mass-to-light ratios.

\begin{center}
\begin{figure*}[htp]
\includegraphics[width=7.5in,viewport=50 50 750 500,clip]{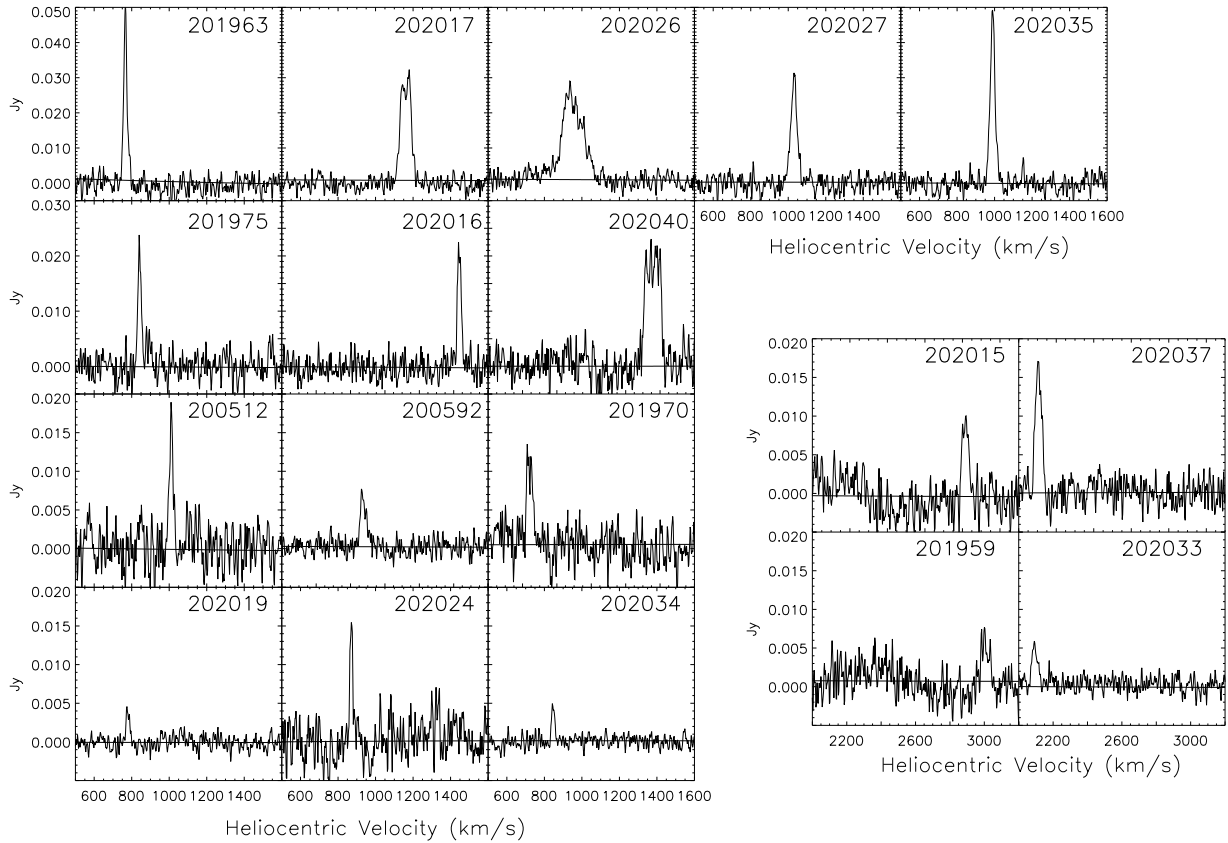}
\caption{Objects successfully detected with the single-pixel L-Band Wide receiver that were originally noted as potential Leo I group members in an optical survey of the region. The values in Table \ref{a1904table} were extracted from these 18 spectra. The y-scales are different for each row. Four of the six background galaxies are plotted separately and over a different velocity range. \label{mosaic}}
\end{figure*}
\end{center}

The results of the targeted single-pixel observations are presented in Table \ref{a1904table}. The first column gives the galaxy's designation in the AGC, while the second column gives the galaxy's designation in KK04. The third and fourth columns show the optical position of the object in J2000 coordinates and the object's B-band magnitude as quoted in KK04. Columns five through nine give the HI parameters measured for each object by the single-pixel observations: the velocity in \kms~(with error), the velocity width in \kms, the integrated line flux in Jy \kms, the rms in mJy, and the signal-to-noise for the detection. For the objects also detected in ALFALFA, the total flux detected by the survey is indicated for comparison.  A mosaic of the spectra obtained for all 18 galaxies is found in Figure \ref{mosaic}.

The last three columns in Table \ref{a1904table} give the derived HI parameters. Suspected members of the M96 group - those objects with 600 \kms~$< v_{helio} < 1200$ \kms~- were placed at 11.1 Mpc. All other distances were determined using the peculiar velocity model as described for Table \ref{paramstable}. HI masses were calculated using these adopted distances for the targeted detections. For the mass-to-light ratios, luminosities were estimated using apparent B-band magnitudes from KK04 and galactic extinction corrections from DIRBE maps \citep{dirbe}. For those objects not found in HI, upper limits are calculated for their HI masses and $M_{HI}/L_{B}$ by placing the objects at 11.1 Mpc and by assuming a peak flux of three times the rms level of each spectrum and a signal width of 50 \kms.  If these galaxies are instead at the adopted Leo II distance of 17.5, the HI mass upper limit should be multiplied by a factor of 2.5. 
In the spectroscopic followup, six of the detected objects were found to be background galaxies and the remaining 12 are considered M96 group members. Of these 12, six detections coincide with the Leo Ring as discussed in Section \ref{ringsec}. Objects not confirmed in the targeted, single-pixel observations either 1) are not actually Leo Group members, 2) lack any HI gas, or 3) contain too little gas to be detected even after several minutes of integration. While the number of objects not detected in ALFALFA but found in the optical survey and vice versa are comparable, all the ALFALFA sources missed by the optical survey were dwarfs. Only a few of the optical galaxies missed by ALFALFA were dwarf-like, and most were L$^*$ galaxies ($m_B \sim 10-12$) lacking gas including the ellipticals NGC~3377 and NGC~3379 and the lenticulars NGC~3384 and NGC~3412. The HI search also has the advantage of automatic redshift information without the need for the time-consuming spectroscopic followup that is required by optical surveys, and particularly important in searches for dwarfs with unclear morphologies. In the case of Leo I, both types of survey were clearly needed to gain a more complete understanding of the group's population, but the blind, HI search proved more successful at finding low-mass group members.

\section{HI Mass Function for Leo I}\label{himfsec}

An important aim of the ALFALFA survey is to determine the HIMF to low HI mass, and eventually to compare how the HIMF might vary in different environments. As noted previously, a drawback of earlier determinations of the HIMF has been the lack of statistics at the low HI mass end, especially in wide area surveys where errors in the distances of the nearby systems which populate the low mass bins are significant. Here, we exploit the group membership to examine the HIMF of the Leo I group alone. Leo I presents an interesting study because of its proximity and because the group also has relatively few bright L$^*$ galaxies compared to other groups \citep{lf}.

\begin{figure}[htp]
\center
\includegraphics[width=3.5in]{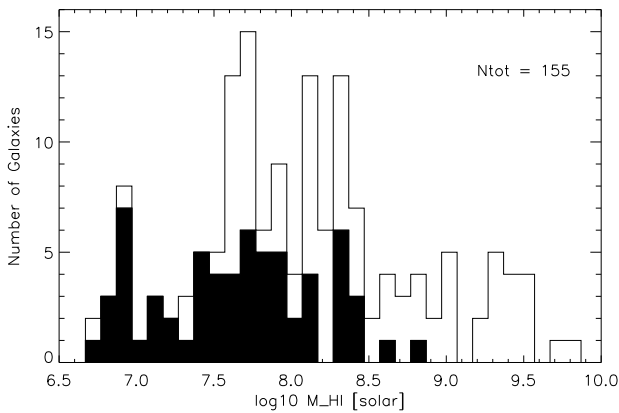}
\caption{Distribution of HI masses for ALFALFA detections in Leo I and in the Leo Cloud. All sources found with ALFALFA from Tables \ref{M96table}, \ref{M66table}, \ref{LeoItable}, and \ref{LIItable} are included. HI flux from all detections constituting the Leo Ring are combined into one entry. Neutral hydrogen from the eight detections making up the Leo Triplet plume is attributed to NGC~3628, and flux from the eight detections constituting the southern clump in the Leo Triplet is added to that of NGC~3627. The distribution peaks at a mass of $10^{7.6} M_{\odot}$. New HI detections are shaded and dominate the low-mass end.\label{mhist}}
\end{figure}

\begin{figure*}[p]
\begin{center}
\includegraphics[width=6in,viewport=25 5 700 500,clip]{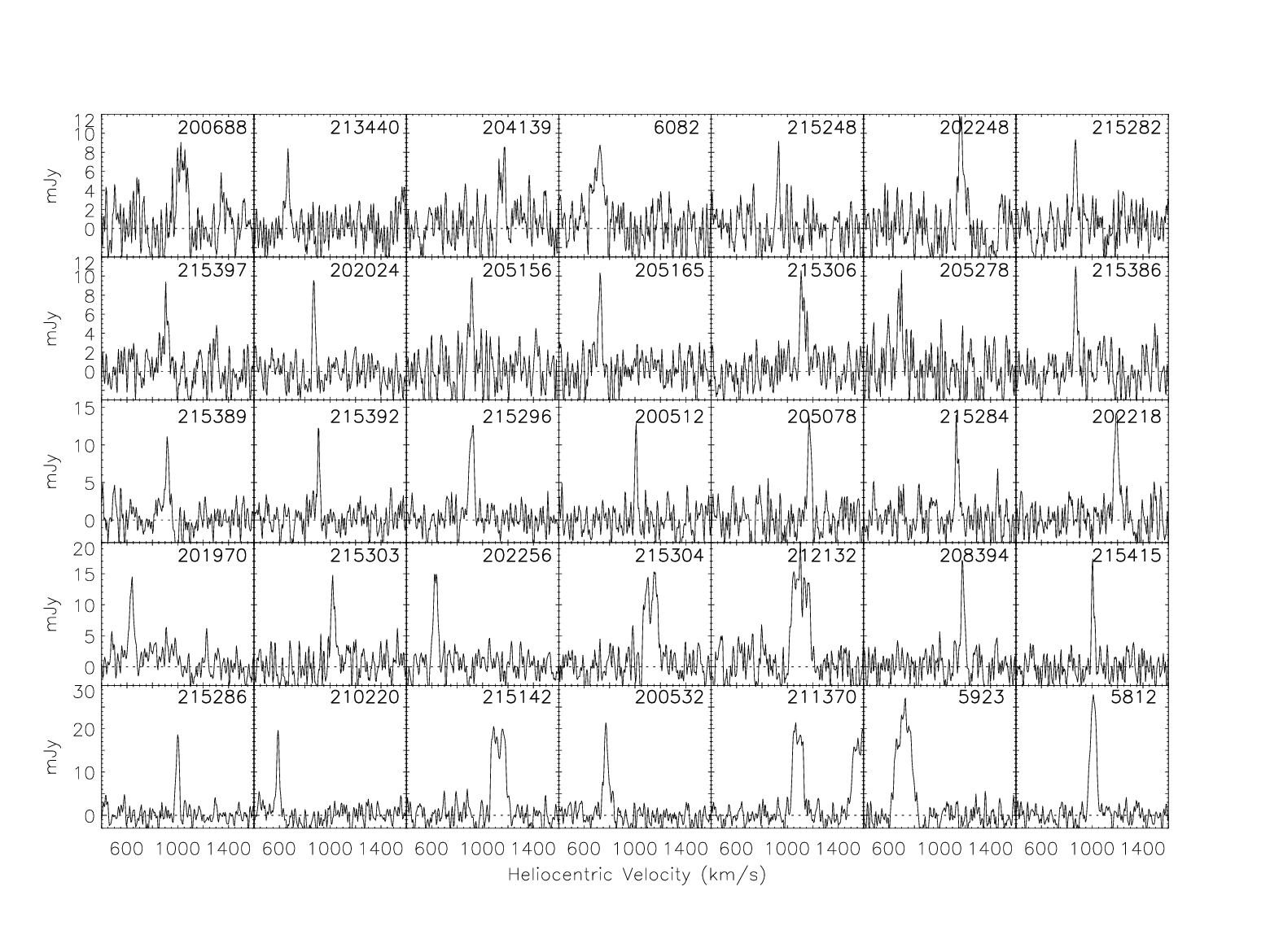}
\includegraphics[width=6in,viewport=18 5 700 500,clip]{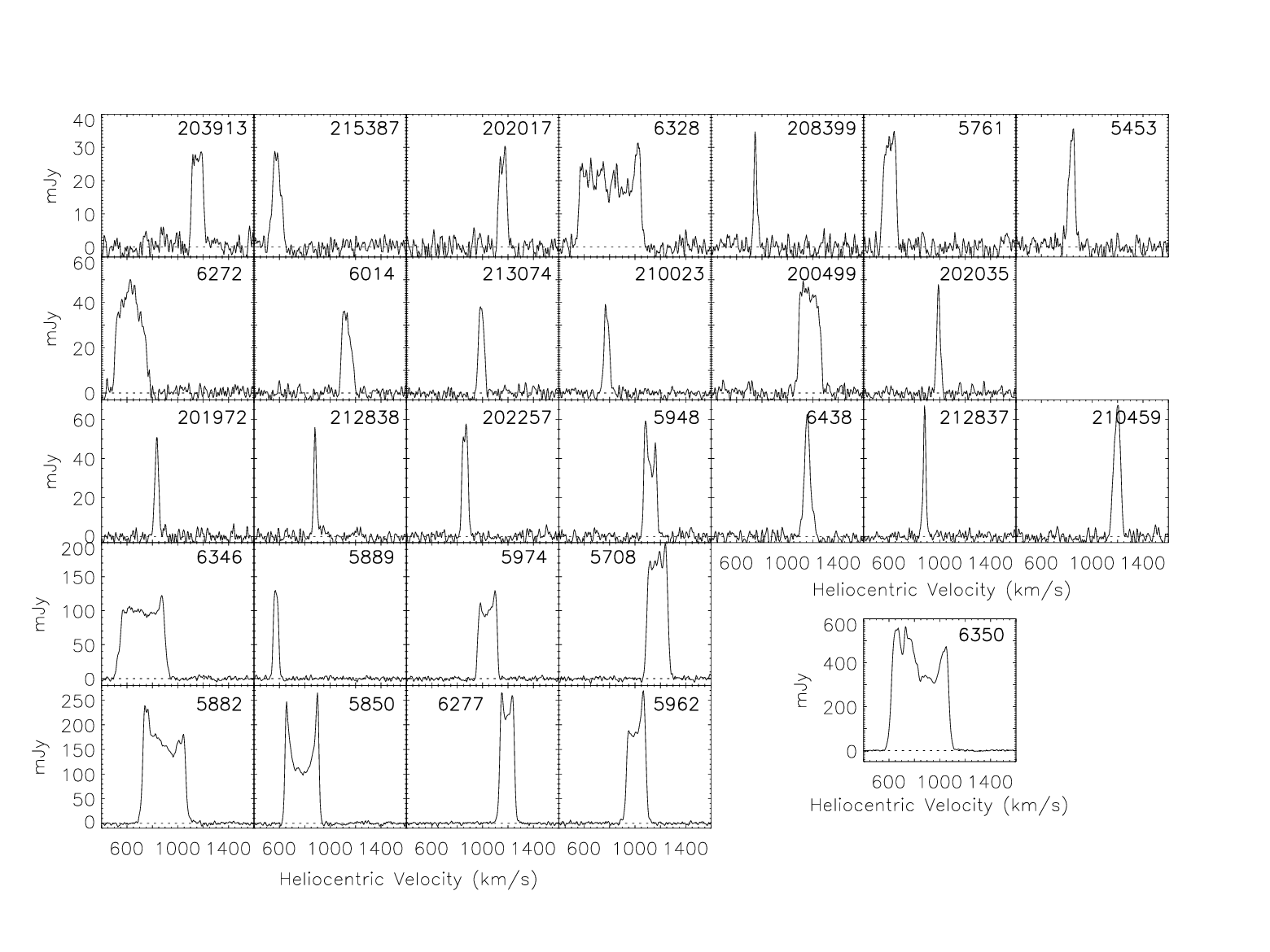}
\caption{The 64 ALFALFA HI spectra of Leo I members used to determine the HIMF in order of increasing peak flux and continued from previous figure. The x-axis is the same for every spectrum, but the y-scales are different for each row. AGC6350 is plotted separately due to its large peak flux value.\label{mosaicappen}}
\end{center}
\end{figure*}

The distribution of HI masses for Leo I and Leo II members found in the available ALFALFA catalog (Tables \ref{M96table}, \ref{M66table}, \ref{LeoItable}, and \ref{LIItable}) is shown in Figure \ref{mhist}. The spread of masses peaks at an HI mass of $10^{7.6} M_{\odot}$ with 91 of the 155 sources having an HI mass of less than $M_{HI} < 10^{8} M_{\odot}$. The ALFALFA sources that are new HI detections are highlighted by the shaded histogram and clearly dominate the low-mass end. In an effort to use a relatively complete and homogeneous sample, we compute an HIMF only from Leo I members detected by the drift scan technique exploited by the ALFALFA survey. Thus, we do not include the galaxies found via single-pixel Arecibo observations, noted in Tables \ref{M96table}, \ref{M66table}, \ref{LeoItable} or any objects from Table \ref{LIItable}. HI flux from all detections constituting the Leo Ring are combined into one data point for determination of the HIMF. Neutral hydrogen from the eight detections making up the Leo Triplet plume is attributed to NGC~3628, and flux from the eight detections constituting the southern clump in the Leo Triplet is added to that of NGC~3627. The resulting dataset includes 65 HI line sources of which 45 have $M_{HI} < 10^{8} M_{\odot}$. Spectra derived from the ALFALFA survey data for all of these sources (with the exception of the Ring) can be found in Figure \ref{mosaicappen}. Calculating a group HIMF from a sample consisting of $69\%$ low-mass objects is a marked improvement on the $1\%$ used for the HIPASS HIMF and almost doubles the $36\%$ used by \cite{Kovacthesis} for the Canes Venatici group HIMF as well as the $34\%$ used by \cite{UMa} in their HIMF for the Ursa Major Cluster. 

\subsection{Completeness Corrections \& Error Estimates }\label{compcorrsec}
As is always the case in determinations of luminosity and mass functions, corrections for completeness are critical. The analysis presented here avoids several problems plaguing determinations of HI mass functions for global samples and even for surveys of larger clusters because Leo I is restricted to a small, nearby volume. In wide area surveys like the \cite{adbshimf} sample of 265 ADBS galaxies, the \cite{s05himf} sample of 2771 optically-selected galaxies, or the \cite{HIPASShimf} sample of 4315 southern HIPASS galaxies, the large search volume results in detection of objects in the foreground of the sample that could not have been detected at larger distances. Each source must then be weighted by $1/V_{max}$ where $V_{max}$ is the maximum volume within which it could have been detected. Even in the blind HI surveys focused on the Ursa Major Cluster \citep{UMa} and on the Canes Venatici Group \citep{Kovacthesis}, the front and back of the cluster are separated enough to require a $1/V_{max}$ correction. All of the Leo I sources, however, are at roughly the same distance, and the minor volume difference between the foreground of the Leo I cloud and the background where Leo I meets Leo II is not large enough to significantly affect the completeness of our sample in terms of volume.

The proximity of Leo I also results in a galaxy sample that is complete down to a lower flux limit throughout the entire survey volume than is possible for wide area surveys probing more distant galaxy populations. At the Leo I distance, the lowest flux reached is 0.20 Jy \kms~which translates to an HI mass of $\sim~10^{6.72} M_{\odot}$. Thus only the lowest mass bin of the Leo I HIMF representing objects with masses between $10^{6.5}$ and $10^{7} M_{\odot}$ needs to be corrected for being populated only down to $10^{6.72} M_{\odot}$. 

As was done for the determinations of the HIMF based on other blind HI surveys \citep{zoa, HIPASShimf, adbshimf}, a correction is needed to account for the dependence of the integrated flux detection limit on the HI line velocity width for ALFALFA detections. As shown in Figure \ref{dotplot}c, an HI spectroscopic survey such as ALFALFA naturally yields a lack of sources with both low fluxes and large velocity widths. To address this bias without over-correcting for an intrinsically small population, we examine the distributions of velocity width for varying HI mass as determined by a complete sample consisting of all of the currently ALFALFA spring sky catalog detections available to us with integrated fluxes greater than 1.0 Jy \kms. It should be noted that although the flux completeness limit dips below 1.0 Jy \kms~to $\sim$.25 Jy \kms~for sources of lower velocity widths, these are not the sources in need of the correction we are seeking here. Since the sample is complete, these distributions reflect the intrinsic nature of the ALFALFA sources without a velocity width-integrated flux selection bias. By comparing these `expected' distributions with those actually observed in Leo I, the deficiency of sources at any given HI mass due to the width-flux selection bias can be corrected. An analysis of the completness of the currently available ALFALFA catalog will be presented in detail by Martin et al. (${\textit{in prep}}$).

Three separate sources of error contribute to the uncertainties in the determination of the HIMF based on the two completeness corrections discussed above and on Poisson statistics. The error associated with the correction for the flux-width dependence is estimated from the errors associated with each fit to the `expected' distributions of HI mass versus velocity width. This source of error does not apply for sources of $M_{HI} > 10^{8.5}M_{\odot}$ since the correction is unneccessary for high masses. The second completeness correction only applies to the lowest mass bin and thus so does the additional associated error. The final contribution of Poisson errors affects all HI mass bins and equals $1/\sqrt(N)$ where N is the number of galaxies in the bin. 

As discussed in \cite{flow}, the use of Hubble flow distances leads to large errors in $M_{HI}$ and the HIMF most significantly for nearby galaxies where peculiar velocites are a more significant fraction of measured recessional velocities. Both \cite{adbshimf} and \cite{HIPASShimf} use flow model distances in their HIMF determinations but claim comparisons with Hubble flow distances show no difference. \cite{HIPASShimf} point out that since peculiar motions only affect a small fraction of their sample, they would not expect a different result between the two methods. However, the population of galaxies most important in determination of the low mass slope are the same galaxies whose distances uncertainties are most impacted by peculiar velocities: the lowest mass objects that make up only a very small fraction of both the HIPASS and ADBS samples. Despite being dominated by these low mass, nearby galaxies, the Leo I sample is able to reduce significantly the errors based on distance measurements by using group distances determined with the help of 10 primary distances. 

\subsection{The Leo I HIMF \& Comparison with other HIMFs}
\begin{figure}[h!]
\begin{center}
\includegraphics[width=3.5in]{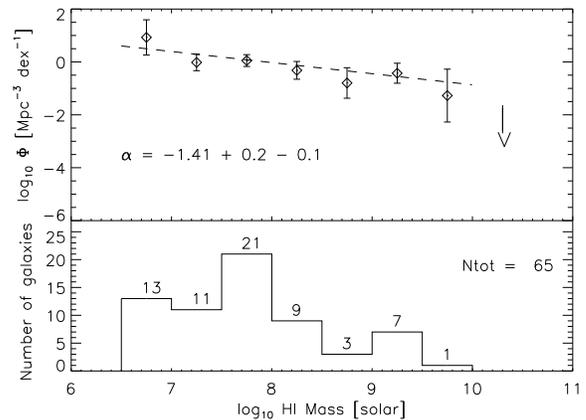}
\caption{The HIMF for Leo I with a linear fit overplotted. The low-mass end slope is well-constrained even given the small sample size due to the large (69$\%$) contribution of low-mass galaxies to the sample. The slope of the linear fit translates to a Schechter function with a low-mass end slope of $\alpha = -1.41 + 0.2 - 0.1$. No objects are found with $M_{HI} > 10^{10}M_{\odot}$. The lack of high-mass galaxies is noted by a downward arrow and suggests that the $\phi^{*}$ and log(M$_{*}$) parameters to the best fit Schechter function should be considered approximations at best.  
\label{himfplot}}
\end{center}
\end{figure}

The Leo I HI mass function is shown in Figure \ref{himfplot}. There are no objects with HI masses greater than $10^{10}M_{\odot}$ in the Leo I volume; we represent this lack of sources with a downward arrow. The current ALFALFA catalog subtends a volume of $\sim~2.5\times 10^6$ Mpc$^3$ and contains 6249 high quality (code 1) detections. 1178 of those have masses of $10^{10}M_{\odot} < M_{HI} < 10^{10.5}M_{\odot}$ and 26 have masses of $10^{10.5}M_{\odot} < M_{HI} < 10^{11}M_{\odot}$.  When compared to the Leo I volume of $\sim$19 Mpc$^3$, the fraction of high mass objects in the larger ALFALFA sample translates to an expected $\sim$0.01 objects of $10^{10}M_{\odot} < M_{HI} < 10^{10.5}M_{\odot}$ and an expected $\sim$0.0002 objects of $10^{10.5}M_{\odot} < M_{HI} < 10^{11}M_{\odot}$ within the volume of Leo I. Thus the lack of objects contributing to the HIMF at the high-mass end do not reflect the lack of such a population but rather the limited volume of our Leo I catalog.

For a Hubble constant of $H_{0} = 70$ \kms~Mpc$^{-1}$, a linear fit to the Leo I HIMF gives a slope of $-$0.41 and a y-offset of 3.33. This slope is identical to that of the best fit Schechter function which has parameters $\phi^{*} = 0.03$ Mpc$^{-3}$, $log_{10}(M_{*}/M_{\odot}) = 10.7$, and $\alpha = -1.41$. The linear fit is overplotted in Figure \ref{himfplot}. Our determination of the Leo I HIMF assumes a volume for the Leo I group of 18.7 Mpc$^3$ with an estimated error of $\pm$ 4 Mpc$^3$. However this error only affects $\phi^{*}$ for the HIMF and does not contribute to errors in $\alpha$. However, the inclusion versus exclusion of Leo I objects that were not clearly in the M96 or M66 groups (as described in Section \ref{groupmemI}) does affect the low mass end slope. If these objects are placed at the M66 group distance of 10.0 Mpc or the M96 group distance of 11.1 Mpc instead of using flow model distances, $\alpha$ is increased by 0.2 or 0.15 respectively. Thus we estimate that for the Leo I HIMF, $\alpha = -1.41 + 0.2 - 0.1$. The values quoted for $\phi^{*}$ and log(M$_*$) are very uncertain given the lack of high mass sources in the Leo I volume and should be considered approximations at best. 

The low mass slope of the Leo I HIMF is compared to six other HIMFs in Table \ref{alphatable}: the \cite{UMa} sample of 32 members of the Ursa Major Cluster (UMa), the \cite{Kovacthesis} survey of 70 objects in the Canes Venatici group (CVn), the \cite{zoa} survey of 2347 objects in the zone of avoidance, the \cite{adbshimf} sample of 265 ADBS galaxies, the \cite{s05himf} sample of 2771 optically-selected galaxies, and the \cite{HIPASShimf} sample of 4315 HIPASS (i.e. HI-selected) galaxies. The mean velocities, as well as the number of total and low-mass galaxies in each sample are quoted. Objects of low HI mass are poorly represented in all but the \cite{Kovacthesis} sample. 

\begin{center}
\begin{deluxetable*}{llrrrc}
\tablewidth{0pt}
\tabletypesize{\tiny} 
\tablecaption{Comparison of the low-mass slope $\alpha$ for different HIMFs \label{alphatable}}
\tablehead{
\colhead{Reference} & \colhead{Sample~~~~~~~~~~~~~~~~~~~~} & \colhead{$\bar{v}$\tablenotemark{*}} & \colhead{$N_{tot}$} & \colhead{~~~~$N_{low}$\tablenotemark{**}} & \colhead{$\alpha$}
\\
 & {Type} & {\kms} & & &
} 
\startdata
this paper & Leo I Group & 931~~ & 65~~ & 45~~~ & -1.41$+$0.20\\
Verheijen \etal ~2001 & Ursa Major Cluster & $\sim$950~~ & 32~~ & 11~~~ & flat \\
Kova\c c 2007\tablenotemark{***} & CVn Group & $\sim$320~~ & 70~~ & 26~~~ & -1.17$\pm$0.08\\
Henning \etal ~2000 & ZOA & 2347~~ & 110~~ & 6~~~ & -1.51$\pm$0.12\\
Rosenberg \& Schneider 2002\tablenotemark{***} & ADBS HI-selected & 3768~~ & 265~~ & 7~~~ & -1.53$\pm$0.12\\
Springob \etal ~2005\tablenotemark{***} & optically-selected & $\sim$5000~~ & 2771~~ & 15~~~ & -1.24$\pm$0.17\\
Zwaan \etal ~2005 & HIPASS HI-selected & 3276~~ & 4315~~ & 44~~~ & -1.37$\pm$0.03\\
\hline
\enddata
\tablenotetext{*}{Mean velocities without a `$\sim$' are calculated directly from the publicly-available data.}
\tablenotetext{**}{Number of objects with $M_{HI} < 10^8 M_{\odot}$.}
\tablenotetext{***}{Errors in $\alpha$ are estimated from the 1$\sigma$ (Kova\c c) or 2$\sigma$ (Springob; Rosenberg \& Schneider) contour presented in the reference.}
\end{deluxetable*}
\end{center}

The slope of -1.41 translates to a prediction of 165 galaxies in Leo I with $M_{HI} < 10^8 M_{\odot}$. To compare the number of low-mass objects expected from the differing HIMFs, the values of $\phi^{*}$ for the global HIMFs (i.e. ZOA, ADBS, Springob \etal~and HIPASS) must be scaled to account for the fact that Leo I is an overdense region. After scaling the $\phi^{*}$ for each global HIMF to match the $\phi^{*}$ of the Leo I HIMF, \cite{s05himf}, \cite{HIPASShimf}, \cite{zoa}, and \cite{adbshimf} expect 3, 9, 25, and 35 HI detections below a mass of $10^{8} M_{\odot}$ respectively. These estimates are all significantly lower than the prediction of 165 low-mass objects from the Leo I HIMF. Without even correcting for completeness, there are more low-mass objects in the Leo I sample than are predicted for all of the global samples. The Leo I sample thus suggests a population of low HI mass objects in the intermediate density group that was not found by earlier, global determinations of the HIMF. 

Since the global HIMFs represent an averaging over a range of galaxy environments and thus are not expected to match the number of low-mass objects found in the overdense Leo I group, a fairer comparison can be made with the HIMF determined for the Canes Venatici group \citep{Kovacthesis}. A quantitative comparison is not made with the Ursa Major HIMF since precise fit parameters are not quoted in that study. After correcting for completeness, \cite{Kovacthesis} expects 61 HI detections below a mass of $10^{8} M_{\odot}$ with a low mass slope of $\alpha = 1.17$. The Leo I and CVn samples are of comparable size, surveyed similar areas on the sky, and both probe nearby groups. (CVn has a group distance of $\sim$4.1 Mpc as determined by primary distances to 17 group members from \cite{CVnLF}.) However, the CVn HIMF is a closer match to the global determinations of the HIMF, particularly the optically-selected sample of \cite{s05himf}. In an optical study of the group, \cite{CVnLF} found the CVn luminosity function also more closely resembled a luminosity function determined for field galaxies than the same function computed for a sample of nearby groups. The CVn group thus may represent an environment where interactions do not play a significant role in galaxy evolution, in contrast to the Leo I environment where strong evidence exists for previous galaxy encounters.

Despite having the highest percentage of low mass objects, the Leo I HIMF does not have the steepest low-mass slope. However, the two determinations of the HIMF that have higher slopes (\cite{zoa} and \cite{adbshimf}) use six and seven galaxies respectively to determine those slopes. These samples also have the lowest number of total galaxies with the exception of those focused on specific galaxy groups. These higher slopes are thus more likely to be poorly constrained due to limitations of sample size rather than representative of a significantly larger population of low-mass objects missed by ALFALFA. The slope derived for the HIMF of \cite{zoa} also most likely suffers from the largest distance errors of the seven samples presented in Table \ref{alphatable} as their distances are based on Hubble flow after correcting measured velocities to the Local Group frame and do not take into account peculiar velocities. 

\subsection{Comparison with Luminosity Functions}
Using a deep optical survey covering a small portion of Leo I aimed at studying the faint-end of the optical luminosity function (LF), (\cite{lf}; hereafter TT02) determined the Leo I LF to have a flat faint end slope and estimated a dwarf-to-giant ratio of 1.6 $\pm$ 0.9, the lowest of the six groups in that study. The TT02 dwarf designation is based on optical properties ($-18 < M_R < -10$) and thus a direct comparison cannot be made to the number of dwarfs presented in the ALFALFA survey's coverage of Leo I. However, as a first order approximation, after labelling all objects with $M_{HI} < 10^8M_{\odot}$ and $W_{50} < 100$ \kms~as ``dwarfs'', the ALFALFA sample, once corrected for completeness, gives a ratio of 9.9 dwarfs for every giant galaxy. The dwarf-to-giant ratio found by ALFALFA does not include lenticular or elliptical galaxies, both of which are found in Leo I and would contribute to the lower ratio found by the TT02 survey. However, given that Leo I has few L$^*$ galaxies compared to other nearby groups, these E/S0s may not make up the whole difference in the optically versus HI-selected ratios. The discrepancy may also be indicative of the existence of a population of gas-rich yet optically faint dwarfs that were either discarded as background galaxies or of too low optical surface brightness to be detected by the optical TT02 survey. TT02 report the ratio of dwarf ellipticals (dEs) to dwarf irregulars (dIs) in Leo I to be 40\%, a much lower fraction than the $\sim$80\% found for all other groups in their study but one, the very poor NGC 1023 group. A comparatively low percentage of dEs would further enhance the contribution made to the Leo I dwarf sample by a population of optically-faint dIs. 

Another factor contributing to the larger number of dwarfs in the HI-selected ALFALFA sample may be that the small portion of the Leo I group covered by the TT02 survey (0.0663 Mpc$^2$ compared to the 12.8 Mpc$^2$ surveyed by ALFALFA) did not accurately represent the group as a whole. Although TT02 probe their search area to fainter magnitudes than previous studies, with their limited strip of coverage TT02 are not even able to include M96 as a group member because the galaxy is far outside their survey bounds. Shallower surveys with larger sky coverage of the region have observed 9 deg$^2$ centered on the core of the M96 group ($\sim$0.3 Mpc$^2$ at a distance of 11.1 Mpc) and found faint end slopes steeper than that for TT02, but still not as steep as for the Leo I HIMF. \cite{fs2} found $\alpha = -1.36$ and \cite{flint} determined $\alpha = -1.17 \pm 0.04$. These samples may not have the sensitivity needed to detect the gas-rich, low surface brightness galaxies more easily found by a blind HI survey of sufficient sensitivity. Alternatively, by focusing only on the center of the M96 group, these optical surveys may be sampling a different dwarf population than that found by the much wider area ALFALFA strategy. Large numbers of gas-rich dIs found further away from the more massive group members would suggest evidence for morphological segregation in Leo I as is seen in the Local Group \citep{grebel}. A complete analysis of the optical properties of the ALFALFA catalog Leo sample will be presented in a future paper to enable direct comparisons with the magnitude limits of these optically-selected samples.

\section{Conclusions}\label{conc}

We presented the results from the ALFALFA survey's current coverage of the Leo region ($09^{h}36^{m} < \alpha < 11^{h}36^{m}$ and $+04^{\circ} < \delta < +16^{\circ}$) which contribute new spectroscopic detections for 48 previously unconfirmed dwarf galaxies with $M < 10^8 M_{\odot}$.

Nearby HI-bearing dwarfs are most commonly characterized by low HI flux and narrow HI line width. In the $\sim$21,000 deg$^2$ of sky covered by HIPASS, the most extensive previous blind HI survey, HIPASS found 44 objects with $M_{HI} < 10^8M_{\odot}$, detected 1 object with an HI flux less than 1 Jy \kms, and made no detections with HI line widths narrower than 30 \kms. In the 354 deg$^2$ surrounding the Leo groups ($\sim 1/60$ of the total coverage of HIPASS), ALFALFA finds 118 objects with $M_{HI} < 10^8M_{\odot}$, detects 260 objects with HI fluxes less than 1 Jy \kms, and makes 55 detections with HI line widths narrower than 30 \kms. After membership determinations are made for the Leo I group, ALFALFA finds 45 low-mass group members, which is more than were detected for the entire southern HIPASS sample.

The HI mass function was determined for the Leo I group, an environment dominated by dwarfs with 69\% of the galaxies in the sample having $M_{HI} < 10^8M_{\odot}$. The best fit Schechter function and linear fits both give the Leo I HIMF a low mass slope of $\alpha = -1.41 + 0.2 - 0.1$. With scaling to account for the higher density environment represented by Leo I, this slope is steeper than that for the optically-selected sample of \cite{s05himf}, the HIPASS survey \citep{HIPASShimf}, and the survey of the Canes Venatici Group done by \cite{Kovacthesis}, but still consistent within the quoted error. Two HIMFs have produced steeper slopes than that for Leo I, the zone of avoidance survey \citep{zoa} and the ADBS, HI-selected survey \citep{adbshimf}, but these surveys have only six and seven total low-mass ($M_{HI} < 10^8M_{\odot}$) detections respectively and most likely carry large distance errors; the low mass slope of the Leo I HIMF was more robustly determined. 

The Leo I HIMF has a steeper low mass end slope than was found for three luminosity functions based on samples of varying depths and sky coverage. In the deepest of these optical surveys designed to find low luminosity dwarfs in Leo I but most limited in sky coverage, \cite{lf} found 1.6 dwarfs for every giant in the group. Using a rough estimate based on HI mass and line width, ALFALFA found a dwarf-to-giant ratio of 9.9, more than six times higher than in the optically-selected sample. This discrepancy may suggest the existence of a population of gas-rich yet optically faint dwarfs not included in the optically-selected sample but is also affected by the lack of E/S0 galaxies in the HI-selected ALFALFA sample. In a direct comparison between an optical survey of the M96 group \citep{KKLeo} and a portion of the ALFALFA survey with the same sky coverage, every group member not found in the optical survey was a dwarf, while only half of the members missed by ALFALFA were dwarfs, and the rest were L$^*$ galaxies (i.e. ellipticals or lenticulars with $m_B\sim 10 - 12$).

The ALFALFA detection statistics in the Leo region reflect the results of the larger survey. In only 20\% of the survey's full intended coverage, ALFALFA has already detected $\sim$300 objects with $M_{HI} < 10^8M_{\odot}$, many of which were previously uncatalogued. The large number of HI-rich dwarfs suggests there may be a significant population of low surface brightness, low-mass galaxies that are missed by optical surveys whether it be due to sensitivity or survey coverage and by HI surveys of lower sensitivity. However, even though the Leo I HIMF is the first HIMF to be dominated by low-mass galaxies, the low-mass slope still falls short of the $\alpha = -1.8$ predicted by cold dark matter simulations. 

The next step to understanding this newly uncovered dwarf population is to examine the optical properties of the ALFALFA dwarf sample. A detailed analysis of the optical properties of Leo I group dwarfs will be presented in a forthcoming paper. A comparison of their surface brightnesses and morphologies with those of the dwarfs uncovered in optical surveys of Leo I will show whether ALFALFA dwarfs were of too low surface brightness to be detected in optical searches or if they resemble background galaxies and thus were systematically excluded from studies of the group. If mass-to-light ratios for these dwarfs are found to be higher than average, these low mass galaxies may represent a dwarf population missed by earlier surveys. Answers to these questions can help determine whether the ALFALFA dwarf population reveals an observational limitation signifcant enough to account for the so-called `missing satellites', or if HI-rich dwarfs alone are not enough.

\acknowledgments
This work was supported by NSF grants AST-0607007 and AST-9397661, and by grants from the Brinson Foundation. We acknowledge the use of NASA's {\it SkyView} facility (http://skyview.gsfc.nasa.gov) located at NASA Goddard Space Flight Center and of the Sloan Digital Sky Survey. Funding for SDSS has been provided by the Alfred P. Sloan Foundation, the Participating Institutions, the National Science Foundation, the US Department of Energy, NASA, the Japanese Monbukagakusho, the Max Planck Society, and the Higher Education Funding Council for England. The SDSS Website is http://www.sdss.org. This research has also made use of the NASA/IPAC Extragalactic Database (NED) which is operated by the Jet Propulsion Laboratory, California Institute of Technology, under contract with the National Aeronautics and Space Administration. IK and VK were partially supported by RFBR grant 07-02-00005.

\clearpage
\begin{landscape}
\begin{deluxetable}{lrcccccrcrcl}
\tablecolumns{12}
\tablewidth{0pt}
\tabletypesize{\scriptsize}
\tablecaption{HI Candidate Detections \label{paramstable}}
\tablehead{
\colhead{Source}  & \colhead{~AGC}   & \colhead{HI Coords (2000)} & \colhead{Opt. Coords (J2000)} & \colhead{~~cz$_{\odot}$}  &
\colhead{$W50 ~(\epsilon_{w})$} & \colhead{~~$F_{c}$} & \colhead{~~S/N} & \colhead{rms} & \colhead{~~Dist}  
 & \colhead{$\log M_{HI}$} & \colhead{Code}     \\
 & {\#~~~} & {hh mm ss.s+dd mm ss} & {hh mm ss.s+dd mm ss} & {~~\kms} & {\kms} & {~~Jy \kms} & & {mJy} & Mpc & {$M_{\odot}$} &
}
\startdata
5-  1  & 192008 & 09 36 03.2 +10 54 08 & 09 36 02.5 +10 54 14 & ~~8518 & 163 (~~9) & ~~0.59 &    4.7 &   2.18 &  126.2 &   ~~9.35 & 2 * \\
5-  2  & 193842 & 09 36 10.0 +11 41 10 & 09 36 08.6 +11 41 21 & ~~8949 & ~38 (~~5) & ~~0.59 &   11.6 &   1.79 &  132.4 &   ~~9.39 & 1   \\
5-  3  & 190385 & 09 36 25.3 +11 20 08 & 09 36 26.0 +11 19 44 & ~~8654 & 338 (~11) & ~~2.99 &   18.2 &   2.00 &  128.2 &  10.06 & 1 * \\
5-  4  & 192364 & 09 36 27.1 +09 36 00 & 09 36 26.8 +09 36 22 & ~~5602 & 121 (~~4) & ~~1.20 &   13.3 &   1.83 &   82.3 &   ~~9.28 & 1   \\
5-  5  & 198344 & 09 36 46.6 +09 02 45 & 09 36 46.4 +09 02 42 & ~~3316 & 106 (~15) & ~~0.62 &    5.8 &   2.31 &   50.4 &   ~~8.57 & 1   \\
5-  6  & 192145 & 09 36 53.0 +11 42 44 & 09 36 53.4 +11 43 01 & ~~8627 & ~49 (~~6) & ~~0.86 &   15.4 &   1.77 &  127.8 &   ~~9.52 & 1 * \\
5-  7  & 191046 & 09 37 00.2 +09 06 37 & 09 37 02.0 +09 06 07 & ~~3058 & ~93 (~11) & ~~0.46 &    5.5 &   1.92 &   46.7 &   ~~8.37 & 2   \\
5-  8  & 198335 & 09 37 00.4 +09 57 54 & 09 37 04.4 +09 57 59 & ~~1517 & ~53 (~~8) & ~~0.37 &    6.5 &   1.73 &   24.2 &   ~~7.71 & 1   \\
5-  9  & 191735 & 09 37 02.4 +09 32 45 & 09 37 02.3 +09 32 24 & ~~5586 & 273 (~14) & ~~1.83 &   12.7 &   1.94 &   82.1 &   ~~9.46 & 1   \\
5- 10  & 192365 & 09 37 09.5 +09 27 49 & 09 37 09.0 +09 27 50 & ~~6719 & 199 (~~4) & ~~2.31 &   21.2 &   1.73 &  100.6 &   ~~9.74 & 1   \\
5- 11  & 192510 & 09 37 24.6 +08 41 42 & 09 37 26.1 +08 41 21 & ~~3308 & ~37 (~20) & ~~0.47 &    6.8 &   2.44 &   50.3 &   ~~8.45 & 1   \\
5- 12  & 191860 & 09 37 56.1 +08 10 45 & 09 37 55.9 +08 10 47 & ~~6201 & ~53 (~~6) & ~~0.49 &    7.6 &   1.95 &   93.2 &   ~~9.00 & 1   \\
5- 13  &   5134 & 09 38 07.3 +09 31 35 & 09 38 07.9 +09 31 23 & ~~3339 & 341 (~~3) & 18.43 &  130.5 &   1.71 &   48.2 &  10.00 & 1   \\
5- 14  & 190408 & 09 38 20.1 +09 26 52 & 09 38 19.3 +09 26 46 & ~~5514 & 175 (~19) & ~~0.87 &    9.0 &   1.62 &   81.1 &   ~~9.13 & 1   \\
5- 15  & 192369 & 09 38 33.9 +09 31 22 & 09 38 32.7 +09 31 16 & ~~5640 & 293 (~44) & ~~0.80 &    5.9 &   1.79 &   82.9 &   ~~9.11 & 2 * \\
5- 16  & 191861 & 09 38 41.1 +08 07 23 & 09 38 40.3 +08 08 10 & ~~3366 & 103 (~10) & ~~1.56 &   17.8 &   1.92 &   51.2 &   ~~8.98 & 1   \\
5- 17  & 193832 & 09 38 48.1 +11 28 26 & 09 38 52.2 +11 29 18 & ~~5883 & ~21 (~~6) & ~~0.34 &    6.7 &   2.35 &   88.6 &   ~~8.80 & 1   \\
5- 18  & 190417 & 09 38 54.5 +09 45 25 & 09 38 53.5 +09 45 01 & ~~5672 & 180 (~16) & ~~1.07 &    8.8 &   2.02 &   83.3 &   ~~9.24 & 1   \\
5- 19  & 192371 & 09 39 14.2 +09 21 54 & 09 39 18.4 +09 22 42 & 14997 & ~44 (~~8) & ~~0.39 &    6.5 &   2.03 &  218.9 & ~~9.64 & 1   \\
5- 20  & 192018 & 09 39 22.6 +10 58 52 & 09 39 23.0 +10 59 13 & 10490 & 193 (~~6) & ~~0.94 &    7.2 &   2.09 &  154.4 & ~~9.72 & 1   \\
\hline
\enddata
\end{deluxetable}
\clearpage
\end{landscape}
\noindent The comments associated with the sources marked with an asterisk in column 12 of Table \ref{paramstable} are given here:\\

\footnotesize
\noindent4-  1: affected by rfi \\
 5-  3: feature bisected by rfi, rfi tentatively interpolated out; params uncertain \\
 5-  6: affected by rfi \\
 5- 15: NGC2939=AGC5134 25 seconds away at 3344 km/s \\
 5- 27: also possible but less likely opt counterpart 1$^{\prime}$ to S \\
 5- 28: on edge of band, ragged data \\
 5- 30: near very strong continuum source \\
 5- 31: blend with U5164 at 094049.7+113306 \\
 5- 32: ambiguous opt id: very blue object at 094105.2+105700 is possible but assigned to 094101.2+105642 because has matching opt cz \\
 5- 39: star superimposed on top of optical image of counterpart \\
 5- 40: disturbed system: blend of U5189, 094253.3+092939 (AGC191298) and 094242.9+092722 (AGC191865) \\
 5- 61: v ragged data \\
 5- 63: optical identification with tiny object is very tentative \\
 5- 65: blue optical galaxy also nearby at 094711.6+100506 (AGC191853) \\
 5- 67: poor centroiding because in region of ragged data \\
 5- 76: equally likely optical counterpart at 094817.2+091044 (AGC712913) \\
 5- 79: blend with optical galaxy at 0948448+105855 (AGC192046) \\
 5- 85: affected by rfi, params uncertain \\
 5- 90: affected by rfi, params uncertain \\
 5-103: equally possible opt id at 095320.3+105539; blend? \\
 5-105: AGC192525 (at 095329.7+083046) also in profile at cz~10740 \\
 5-110: opt id is tentative - HI centroid is off opt position by more than 1 arcmin (seems high for this s/n) \\
 5-117: HI emission blended with IC578 (AGC5337) \\
 5-124: v ragged data \\
 5-141: crowded optical field, opt id somewhat ambiguous \\
 5-156: equally likely optical counterpart at 100435.0+102424 (CGCG 064-049/AGC200031); blend? \\
 5-157: also possible but less likely opt counterpart at 100442.5+102251 (AGC202289) \\
 5-162: HI looks to be a blend of CGCG 064-053 (AGC200042) and opt galaxy at 1005259+114253 (AGC205066) \\
 5-186: no identifiable opt counterpart, uncertain separation from MW HI, HVC \\
 5-188: no identifiable opt counterpart, but s/n marginal \\
 5-196: affected by rfi, params very uncertain \\
 5-203: possibly affected by rfi; params somewhat uncertain \\
 5-206: no identifiable optical counterpart \\
 5-227: alternate opt id at 1024372+101650 (205096), assigned to 202371 because has matching opt cz \\
 5-233: affected by rfi, params very uncertain \\
 5-238: possible but less likely opt counterpart at 102639.7+105710 (AGC200380) \\
 5-240: HI emission might be blended with that of opt galaxy at 102656.2+080908 (AGC201443) which is nearby and at similar cz \\
 5-241: no identifiable optical counterpart, extended HVC \\
 5-259: blend with emission of AGC202558 but recovery of this signal better than fair \\
 5-260: part of crowded group in field, blend mainly with AGC200456 at 103310.3+115326 and AGC200466 at 103333.4+115217 \\
 5-263: part of crowded group in field, severely nasty blend with AGC200463 at 103332.6+115232, putative contribution from AGC200463 tentatively interpolated out; params very uncertain \\
 5-264: part of crowded group in field, contact pair with AGC200466 at 103333.4+115217, severely nasty blend; params very uncertain \\
 5-265: tight blend with AGC200466 and AGC200463 \\
 5-266: part of crowded group in field, blend mainly with AGC200466 at 103333.4+115217; params very uncertain \\
 5-268: affected by rfi; params very uncertain \\
 5-276: equally likely opt counterpart at 103656.3+104015 (AGC202411); HI looks to be a blend of the two \\
 5-281: blend with CGCG 065-071=AGC200495 at 103721.7+094615 \\
 5-282: HI blended with that of opt gal at 103715.7+094936 (AGC202227) \\
 5-283: on edge of bandpass, ragged data; params uncertain \\
 5-286: affected by rfi; params uncertain \\
 5-287: affected by rfi; params very uncertain \\
 5-291: affected by rfi; params very uncertain \\
 5-293: HVC extends over 15' \\
 5-309: extended HI emission \\
 5-316: affected by rfi, possible blend with AGC205470 at 104531.8+082923; params very uncertain \\
 5-321: extended HI emission \\
 5-323: part of the Leo Ring \\
 5-327: blend with emission of AGC202645 \\
 5-334: ambiguous opt id; also possible at 104956.9+090400 \\
 5-342: possible blend with emission of gal at 105102.0+113734 (AGC202660) \\
 5-352: HI emission blends with that of N3444 (AGC6004) \\
 5-356: HI emission blends with gal at 105237.4+080024 (AGC202236) \\
 5-362: possible but less likely opt counterpart at 105333.4+083834 \\
 5-363: affected by rfi; params very uncertain \\
 5-370: blend with emission of AGC208359 at 105508.7+094936; params probably affected \\
 5-371: ambiguous opt id; HI probably a blend \\
 5-372: blend with emission of AGC205333 at 105456.6+095235; params probably affected \\
 5-378: blended with AGC201030 at 105603.9+094422; params uncertain \\
 5-386: alternate (equally likely) opt id at 105734.1+091037 \\
 5-394: poorly determined position and width \\
 5-397: affected by rfi; params uncertain \\
 5-402: affected by rfi; params uncertain \\
 5-406: affected by rfi; params uncertain \\
 5-415: affected by rfi; params mildly uncertain \\
 5-420: affected by rfi; params very uncertain \\
 5-432: HI emission a heavy blend of AGC213587 at 110352.1+090948 and AGC213586 at 110351.4+090759 \\
 5-436: affected by rfi; params mildly uncertain \\
 5-444: ambiguous opt id: also possible at 110619.2+082856; blend? \\
 5-449: part of triple system, mild blend with AGC210072=CGCG 067-009 at 110819.6+100224 and AGC213590 at 110830.7+095519 \\
 5-450: part of triple system, mild blend with AGC210073=CGCG 067-008 at 110819.6+095702 and AGC213590 at 110830.7+095519; params uncertain \\
 5-453: part of triple system, mild blend with AGC210073=CGCG 067-008 at 110819.6+095702 and AGC210072=CGCG 067-009 at 110819.6+100224 \\
 5-462: affected by rfi; params very uncertain \\
 5-477: affected by rfi; params very uncertain \\
 5-478: large HI/opt position offset may be significant \\
 5-486: for feature assigned opt id, SDSS gives unrealistic z=2.2 \\
 5-487: alternate opt id at 111728.5+114442 \\
 5-488: ambiguous opt id; also possible at 111736.5+084646 (AGC213692) and at 111743.2+084634 (AGC213693); blend? \\
 5-490: no identifiable optical counterpart \\
 5-493: affected by rfi; params uncertain \\
 5-498: ambiguous opt id; also possible at 112137.4+114801; blend? \\
 5-499: ambiguous opt id; also possible at 112151.48+115325.1 \\
 5-500: HI source in v close pair: ambiguous opt id; also possible at 112157.6+102956; blend? \\
 5-507: ambiguous opt id; also possible at 112514.2+113148; blend? \\
 5-520: part of crowded group in field, blend mainly with AGC210354 at 112812.9+090344, AGC6477 at 112840.0+090555, and AGC6475 at 112831.0+090614; params uncertain \\
 5-521: part of crowded group in field, blend mainly with AGC6470 at 112814.8+090848, AGC6477 at 112840.0+090555, and AGC6475 at 112831.0+090614; params uncertain \\
 5-523: part of crowded group in field, severely nasty blend with AGC6477 at 112840.0+090555, putative contribution from AGC6477 tentatively interpolated out; params very uncertain \\
 5-525: part of crowded group in field, severely nasty blend with AGC6475 at 112831.0+090616; params uncertain \\
 5-526: part of crowded group in field, severely nasty blend with AGC6482 at 112903.8+090641; params very uncertain \\
 5-527: part of crowded group in field, severely nasty blend with AGC210368 at 112900.5+090522, putative contribution from AGC210368 tentatively interpolated out; params very uncertain \\
 5-534: extended HI \\
 5-538: ambiguous opt id; also possible at 113244.8+082544, params affected by rfi \\
 5-540: evidence for extended HI \\
 5-543: alternate opt id at 113435.7+081455 \\
\clearpage

\LongTables
\begin{deluxetable}{llcccccc}
\tablewidth{0pt}
\tabletypesize{\scriptsize} 
\tablecaption{Probable Leo Cloud Members \label{LIItable}}
\tablehead{
\colhead{AGC} & \colhead{Other\tablenotemark{a}} & \colhead{Opt Position} & \colhead{HI cz$_{\odot}$} & \colhead{W50} & \colhead{$F_c$} &
\colhead{Dist\tablenotemark{b}} & \colhead{$\log M_{HI}$} 
     \\
{\#} & {~Name} & {(J2000)} & {\kms} & {\kms} & {Jy \kms} & {Mpc} & {$M_{\odot}$} 
} 
\startdata
198335 &         & 09 37 04.4 +09 57 59 & 1517 &  ~~53 &   0.37 & 24.2 & 7.66 \\
192830 &         & 09 39 22.3 +04 57 08 & 1886 & 167 &   3.22 & 29.5 & 8.82 \\
192937 &         & 09 40 21.1 +04 44 06 & 1983 &  ~~44 &   0.29 & 30.9 & 7.75 \\
192833 &         & 09 40 56.3 +05 02 41 & 1871 &  ~~49 &   1.21 & 29.3 & 8.40 \\
193813 &         & 09 42 50.9 +04 53 24 & 1939 &  ~~87 &   0.63 & 30.3 & 8.06 \\
198337 &         & 09 42 51.2 +09 38 00 & 1461 &  ~~34 &   0.62 & 23.4 & 7.90 \\
192835 &         & 09 43 02.2 +05 01 45 & 1963 &  ~~95 &   1.32 & 30.6 & 8.44 \\
191849 &         & 09 44 37.1 +10 00 46 & 1483 &  ~~62 &   1.94 & 23.7 & 8.40 \\
191869 &         & 09 44 58.9 +08 22 12 & 1733 & 163 &   4.22 & 27.3 & 8.86 \\
198456 &         & 09 46 42.4 +07 08 07 & 1886 &  ~~57 &   0.62 & 29.5 & 8.10 \\
193921 &         & 09 49 14.9 +15 48 27 & 1449 &  ~~39 &   0.61 & 23.3 & 7.80 \\
  5271 & N3020   & 09 50 06.7 +12 48 46 & 1438 & 217 &  31.65~~ & 21.9\tablenotemark{*} & 9.55 \\
  5275 & N3024   & 09 50 27.2 +12 45 55 & 1418 & 246 &  26.82~~ & 22.8 & 9.51 \\
192239 &         & 09 50 36.3 +12 48 33 & ~~~~~~~1335 (opt) & ~~... & ... & 21.6 & ... \\
192423 &         & 09 54 30.5 +09 52 12 & 1488 &  ~~40 &   0.45 & 23.8 & 7.77 \\
192959 &         & 09 54 35.7 +04 23 08 & 1774 &  ~~77 &   0.98 & 27.8 & 8.22 \\
  5325 & N3049   & 09 54 49.7 +09 16 16 & 1497 & 203 &  11.55~~ & 15.3\tablenotemark{*} & 8.80 \\
  5328 & N3055   & 09 55 18.0 +04 16 12 & 1821 & 266 &  11.26~~ & 28.5 & 9.33 \\
190600 & 063-105 & 09 55 29.3 +08 23 27 & 1281 & 101 &   2.83 & 20.7 & 8.45 \\
192766 &         & 09 57 21.1 +06 25 03 & ~~~~~~~1665 (opt) & ~~... & ... & 26.3 & ... \\
192960 &         & 09 55 37.8 +04 28 36 & 1942 &  ~~61 &   0.77 & 30.3 & 8.15 \\
205283 & HIonly  & 10 01 30.9 +13 21 35 & 1954 &  ~~69 &   0.54 & 30.6 & 7.99 \\
204045 &         & 10 02 00.0 +04 47 27 & ~~~~~~~1693 (opt) & ~~... & ... & 26.7 & ... \\
200879 & 036-027 & 10 04 08.7 +06 30 38 & 1263 &  ~~42 &   0.62 & 20.4 & 7.86 \\
202297 &         & 10 06 03.8 +10 38 16 & 1565 & 258 &   1.88 & 25.0 & 8.43 \\
205108 &         & 10 06 40.3 +12 19 00 & 1487 &  ~~26 &   0.55 & 23.9 & 7.89 \\
203862 &         & 10 07 04.5 +05 00 25 & 1739 &  ~~34 &   0.90 & 27.4 & 8.16 \\
203863 &         & 10 07 24.1 +05 19 31 & ~~~~~~~1603 (opt) & ~~... & ... & 25.4 & ... \\
205076 & FGC120A & 10 09 17.4 +05 24 15 & 1701 &  ~~83 &   1.02 & 26.8 & 8.21 \\
203432 &         & 10 10 20.6 +07 45 13 & ~~~~~~~1268 (opt) & ~~... & ... & 20.6 & ... \\
  5504 & 036-059 & 10 12 49.0 +07 06 11 & 1545 & 147 &   3.91 & 24.7 & 8.74 \\
  5522 & 036-065 & 10 13 59.0 +07 01 24 & 1218 & 211 &  34.25~~ & 19.8 & 9.50 \\
201993 & KKH 60  & 10 15 59.4 +06 48 16 & 1620 &  ~~94 &   1.53 & 25.8 & 8.37 \\
202131 &         & 10 17 09.2 +04 20 43 & ~~~~~~~1308 (opt) & ~~... & ... & 21.1 & ... \\
  5551 &         & 10 17 11.8 +04 19 50 & 1344 &  ~~56 &   4.49 & 21.7 & 8.69 \\
208392 &         & 10 18 03.7 +04 18 35 & 1322 &  ~~34 &   0.53 & 21.3 & 7.75 \\
  5633 & 094-035 & 10 24 40.0 +14 45 23 & 1382 & 167 &  15.63~~ & 22.6 & 9.27 \\
  5646 & 094-048 & 10 25 53.0 +14 21 48 & 1368 & 221 &   9.49 & 22.4 & 9.05 \\
208295 &         & 10 28 27.2 +08 10 26 & 1491 &  ~~91 &   1.06 & 24.1 & 8.14 \\
204135 &         & 10 31 37.3 +04 34 22 & ~~~~~~~1202 (opt) & ~~... & ... & 19.6 & ... \\
202244 &         & 10 31 40.8 +13 50 04 & 1288 & 102 &   1.89 & 21.3 & 8.30 \\
202016\tablenotemark{c} &         & 10 33 19.2 +10 11 22 & 1433 &  ~~32 &   0.57 & 23.3 & 7.82 \\
205161 &         & 10 34 05.6 +15 46 50 & 1218 & 114 &   1.03 & 20.3 & 8.00 \\
  5741 & I622    & 10 34 42.8 +11 11 48 & 1389 & 347 &   3.76 & 22.8 & 8.63 \\
202262 & FGC125a & 10 37 28.7 +12 23 46 & 1330 &  ~~59 &   1.81 & 22.0 & 8.31 \\
203080 &         & 10 41 41.0 +13 49 30 & ~~~~~~~1271 (opt) & ~~... & ... & 17.5\tablenotemark{**} & ... \\
  5826 & N3338   & 10 42 07.6 +13 44 48 & 1298 & 339 &  91.69~~ & 17.5\tablenotemark{**} & 9.82 \\
203082 &         & 10 42 26.5 +13 57 26 & 1277 &  ~~41 &   0.52 & 17.5\tablenotemark{**} & 7.58 \\
  5832 & 065-089 & 10 42 48.6 +13 27 35 & 1217 & 102 &   5.47 & 17.5\tablenotemark{**} & 8.59 \\
200543 & 065-090 & 10 43 05.5 +13 30 42 & 1256 &  ~~70 &   2.98 & 17.5\tablenotemark{**} & 8.29 \\
  5842 & N3346   & 10 43 38.9 +14 52 16 & 1258 & 162 &  15.20~~ & 17.5\tablenotemark{**} & 9.03 \\
200552\tablenotemark{d} &         & 10 43 57.0 +13 23 14 & 1210 &  ~~99 &   1.51 & 17.5\tablenotemark{**} & 8.04 \\
205270 &         & 10 45 09.8 +15 26 59 & 1230 &  ~~51 &   0.42 & 17.5\tablenotemark{**} & 7.43 \\
  5914 & N3389   & 10 48 28.6 +12 31 57 & 1301 & 266 &  21.89~~ & 21.4\tablenotemark{*} & 9.37 \\
200598 & 066-025 & 10 48 56.8 +12 11 40 & 1321 & 125 &   4.21 & 17.5\tablenotemark{**} & 8.47 \\
200600 & 066-024 & 10 48 59.7 +10 50 07 & 1939 & 120 &   1.33 & 17.5\tablenotemark{**} & 7.97 \\
205309 & HIonly  & 10 49 07.6 +12 22 34 & 1342 &  ~~33 &   1.40 & 17.5\tablenotemark{**} & 8.01 \\
205310 & HIonly  & 10 49 11.5 +12 29 39 & 1379 &  ~~50 &   3.41 & 17.5\tablenotemark{**} & 8.39 \\
200603 & 066-029 & 10 49 17.1 +12 25 20 & 1376 &  ~~68 &   3.75 & 17.5\tablenotemark{**} & 8.43 \\
202253 &         & 10 49 26.7 +12 15 28 & ~~~~~~~1319 (opt) & ~~... & ... & 17.5\tablenotemark{**} & ... \\
205197 &         & 10 49 42.8 +13 49 41 & 1332 &  ~~42 &   0.38 & 17.5\tablenotemark{**} & 7.38 \\
205198 &         & 10 50 01.8 +13 47 05 & 1322 &  ~~53 &   0.62 & 17.5\tablenotemark{**} & 7.71 \\
202260 & F640V02 & 10 57 38.2 +13 58 42 & 1238 &  ~~92 &   2.72 & 17.5\tablenotemark{**} & 8.29 \\
  6077 & N3485   & 11 00 02.4 +14 50 28 & 1432 & 135 &  21.66~~ & 17.5\tablenotemark{**} & 9.19 \\
202040\tablenotemark{c} & LeG35   & 11 03 02.0 +08 02 53 & 1359 &  ~~96 &   1.74 & 17.5\tablenotemark{**} & 8.10 \\
219117 &         & 11 03 46.7 +08 34 19 & 1738 &  ~~68 &   0.65 & 17.5\tablenotemark{**} & 7.66 \\
213757 &         & 11 05 59.6 +07 22 25 & 1640 &  ~~57 &   0.58 & 17.5\tablenotemark{**} & 7.57 \\
  6158 & N3524   & 11 06 32.1 +11 23 06 & ~~~~~~~1321 (opt) & ~~... & ... & 17.5\tablenotemark{**} & ... \\
215262 &         & 11 06 35.3 +12 13 48 & 1606 &  ~~63 &   0.55 & 17.5\tablenotemark{**} & 7.56 \\
  6167 & N3526   & 11 06 56.8 +07 10 26 & 1416 & 196 &   8.96 & 19.8\tablenotemark{*} & 8.92 \\
  6169 & 066-115 & 11 07 03.4 +12 03 34 & 1551 & 241 &   9.80 & 17.5\tablenotemark{**} & 8.84 \\
210082 & 067-014 & 11 09 23.2 +10 50 03 & 1555 &  ~~66 &   2.46 & 17.5\tablenotemark{**} & 8.26 \\
  6209 & N3547   & 11 09 55.9 +10 43 12 & 1584 & 204 &   7.42 & 18.1\tablenotemark{*} & 8.74 \\
210111 & 067-022 & 11 10 25.1 +10 07 34 & 1320 &  ~~60 &   2.72 & 17.5\tablenotemark{**} & 8.29 \\
213064 &         & 11 10 54.5 +09 37 19 & 1604 & 124 &   3.26 & 17.5\tablenotemark{**} & 8.36 \\
  6233 & 039-056 & 11 11 28.3 +06 54 26 & 1605 & 212 &   1.82 & 26.0 & 8.45 \\
  6245 & I676    & 11 12 39.8 +09 03 21 & 1421 & 177 &   1.29 & 17.5\tablenotemark{**} & 7.94 \\
  6248 &         & 11 12 51.7 +10 12 00 & 1286 &  ~~26 &   2.29 & 17.5\tablenotemark{**} & 8.21 \\
213796 &         & 11 12 52.7 +07 55 19 & 1412 &  ~~78 &   0.55 & 17.5\tablenotemark{**} & 7.57 \\
212097 & 039-068 & 11 13 00.1 +07 51 43 & 1396 & 118 &   2.01 & 17.5\tablenotemark{**} & 8.16 \\
215280 &         & 11 13 16.3 +15 24 28 & 1479 &  ~~93 &   0.84 & 17.5\tablenotemark{**} & 7.78 \\
215240 &         & 11 13 50.8 +09 57 39 & 1610 &  ~~34 &   0.45 & 17.5\tablenotemark{**} & 7.49 \\
219197 &         & 11 13 55.2 +04 06 19 & 1609 &  ~~63 &   0.88 & 25.9 & 8.14 \\
215186 &         & 11 17 01.2 +04 39 44 & 1455 &  ~~66 &   0.27 & 24.0 & 7.58 \\
215241 &         & 11 17 02.7 +10 08 36 & 1765 & 120 &   1.80 & 17.5\tablenotemark{**} & 8.11 \\
  6306 &         & 11 17 27.4 +04 36 16 & 1746 & 108 &   4.84 & 27.8 & 8.94 \\
  6305 & N3611   & 11 17 30.0 +04 33 19 & 1612 & 375 &  14.06~~ & 26.0 & 9.29 \\
215287 &         & 11 19 45.1 +15 30 08 & 1334 & 103 &   0.73 & 17.5\tablenotemark{**} & 7.72 \\
214314 &         & 11 22 11.1 +04 39 42 & ~~~~~~~1305 (opt) & ~~... & ... & 22.1 & ... \\
  6387 & I2763   & 11 22 18.1 +13 03 53 & 1572 & 132 &   2.90 & 16.6\tablenotemark{*} & 8.27 \\
213511 &         & 11 22 23.4 +11 47 38 & 1571 &  ~~61 &   0.40 & 17.5\tablenotemark{**} & 7.44 \\
219201 &         & 11 22 31.4 +05 31 29 & 1575 &  ~~24 &   0.35 & 25.7 & 7.72 \\
213512 & I2781   & 11 22 50.7 +12 20 41 & 1544 &  ~~72 &   1.16 & 17.5\tablenotemark{**} & 7.95 \\
215290 &         & 11 22 59.1 +12 27 38 & 1613 &  ~~42 &   0.97 & 17.5\tablenotemark{**} & 7.85 \\
  6420 &         & 11 24 26.2 +11 20 30 & 1059 & 255 &  39.28~~ & 16.3\tablenotemark{*} & 9.39 \\
214317 &         & 11 25 05.4 +04 07 16 & 1619 & 130 &   2.36 & 26.2 & 8.59 \\
214318 &         & 11 25 40.0 +04 40 36 & 1527 & 123 &   0.46 & 25.1 & 7.65 \\
219119 &         & 11 26 03.4 +08 04 32 & 1567 &  ~~35 &   0.44 & 25.8 & 7.79 \\
214319 &         & 11 26 08.3 +04 03 45 & 1525 &  ~~49 &   0.81 & 25.0 & 8.09 \\
219202 &         & 11 27 10.9 +05 08 56 & 1518 &  ~~70 &   0.69 & 25.0 & 7.93 \\
219203 &         & 11 27 28.9 +05 37 02 & 1512 &  ~~28 &   0.32 & 25.0 & 7.66 \\
  6474 & N3692   & 11 28 24.0 +09 24 26 & 1716 & 408 &  10.55~~ & 27.8 & 9.28 \\
213939 &         & 11 28 24.3 +06 07 04 & 1571 &  ~~45 &   1.02 & 15.8 & 8.20 \\
  6498 & N3705   & 11 30 07.6 +09 16 36 & 1019 & 345 &  41.38~~ & 17.2\tablenotemark{*} & 9.46 \\
213169 &         & 11 35 18.4 +04 57 17 & 1417 &  ~~37 &   0.90 & 23.9 & 8.08 \\
\hline
\enddata
\tablenotetext{a}{Positions indicate the centroid of the optical counterpart unless the object is noted as an HI-only detection, in which case the position represents the centroid of the HI.}
\tablenotetext{b}{Objects are given flow model distances except when a secondary distance is known (marked by *) or a group distance was assigned (marked by **).}
\tablenotetext{c}{HI parameters come from single pixel results (see Section \ref{optselect})}
\tablenotetext{d}{HI parameters come from previously catalogued single-pixel Arecibo observations. See the HI archive for details.}
\end{deluxetable}
\clearpage

\begin{landscape}
\begin{deluxetable}{clclrclcrcrr}
\tablewidth{0pt}
\tabletypesize{\tiny} 
\tablecaption{Results of Single Pixel, Targeted HI Observations \label{a1904table}}
\tablehead{
\colhead{AGC} & \colhead{Other} & \colhead{Opt Position} & \colhead{m$_B$} & \colhead{HI cz$_{\odot} ~(\epsilon_{cz})$} &
\colhead{$W50$} & \colhead{$F_c ~[F_{c,ALF}]$} & \colhead{rms} & \colhead{~S/N} & \colhead{Dist} & \colhead{$\log M_{HI}$} &
\colhead{$\log M_{HI}/L_{B}$}
     \\
{\#} & {~~Name} & {(J2000)} & & {\kms~~~~} & {\kms} & {~~~~Jy \kms} & {mJy} & & {Mpc} & {$M_{\odot}$~~~~} & {$M_{\odot}/L_{\odot}$~~~}
} 
\startdata
202015 & LeG01 & 10 31 53.8 +12 55 34 & 18.7 & 2815 (2)~~~~ & ~42 & ~~~~0.36 [0.42] & 1.7 & 10.1 & 43.1 & 8.20~~~~ & 2.16~~~~~~~~\\
202016 & LeG02 & 10 33 19.2 +10 11 20 & 19.1 & 1433 (1)~~~~ & ~28 & ~~~~0.57 [0.59] & 2.0 & 17.3 & 17.5 & 7.61~~~~ & 2.52~~~~~~~~\\
202017 & LeG03 & 10 35 48.9 +08 28 49 & 17.8 &  1160 (2)~~~~ & ~68 & ~~~~1.97 [1.93] & 1.7 & 45.1 & 11.1 & 7.76~~~~ & 2.55~~~~~~~~\\
202018 & LeG04 & 10 39 40.2 +12 44 05 & 18.7 & -~~~~~~~~~ & - & ~~~~~~~~- & 1.9 & - & - & $<$ 6.92~~~~ & $<$ 2.07~~~~~~~~\\
202019 & FS 01 & 10 39 43.3 +12 38 03 & 16.77 & ~~780 (3)~~~~ & ~22 & ~~~~0.08 & 0.8 & 6.6 & 11.1 & 6.37~~~~ & 0.75~~~~~~~~\\
200512 & P031727 & 10 39 55.6 +13 54 34 & 18.3 & 1011 (2)~~~~ & ~20 & ~~~~0.42 [0.30] & 2.4 & 13.2 & 11.1 & 7.09~~~~ & 2.07~~~~~~~~\\
202020 & LeG09 & 10 42 34.6 +12 09 01 & 18.5 & -~~~~~~~~~ & - & ~~~~~~~~- & 2.5 & - & - & $<$ 7.04~~~~ & $<$ 2.11~~~~~~~~\\
202021 & LeG10 & 10 43 55.4 +12 08 06 & 19.2 & -~~~~~~~~~ & - & ~~~~~~~~- & 1.7 & - & - & $<$ 6.86~~~~ & $<$ 2.21~~~~~~~~\\
202022 & LeG11 & 10 44 02.1 +15 35 19 & 18.8 & -~~~~~~~~~ & - & ~~~~~~~~- & 1.8 & - & - & $<$ 6.90~~~~ & $<$ 2.08~~~~~~~~\\
202023 & LeG12 & 10 44 07.3 +11 31 58 & 19.1 & -~~~~~~~~~ & - & ~~~~~~~~- & 2.0 & - & - & $<$ 6.94~~~~ & $<$ 2.24~~~~~~~~\\
202024 & FS 09 & 10 44 57.3 +11 55 01 & 17.43 & ~~870 (2)~~~~ & ~18 & ~~~~0.31 [0.23] & 2.5 & 9.2 & 11.1 & 6.95~~~~ & 1.60~~~~~~~~\\
202025 & FS 13 & 10 46 14.4 +12 57 35 & 18.7 & -~~~~~~~~~ & - & ~~~~~~~~- & 0.9 & - & - & $<$ 6.58~~~~ & $<$ 1.73~~~~~~~~\\
201990 & FS 14 & 10 46 24.7 +14 01 26 & 18.3 & -~~~~~~~~~ & - & ~~~~~~~~- & 1.6 & - & - & $<$ 6.84~~~~ & $<$ 1.82~~~~~~~~\\
202026\tablenotemark{a} & FS 15 & 10 46 30.0 +11 45 20 & 19.0 & ~~954 (7)~~~~ & 126 & ~~~~3.24 & 1.8 & 49.3 & 11.1 & 7.97~~~~ & 3.24~~~~~~~~\\
202027\tablenotemark{b} & FS 17 & 10 46 41.3 +12 19 37 & 16.98 & 1030 (2)~~~~ & ~37 & ~~~~1.24 & 2.1 & 30.9 & 11.1 & 7.56~~~~ & 2.03~~~~~~~~\\
201970\tablenotemark{c} & LeG18 & 10 46 52.2 +12 44 39 & 18.9 & ~~643 (2)~~~~ & ~43 & ~~~~0.49 [0.61] & 2.1 & 11.1 & 11.1 &
7.15~~~~ & 2.39~~~~~~~~\\
201971 & FS 20 & 10 46 54.8 +12 47 16 & 18.2 & -~~~~~~~~~ & - & ~~~~~~~~- & 1.7 & - & - & $<$ 6.86~~~~ & $<$ 1.81~~~~~~~~\\
201975\tablenotemark{a} & LeG21 & 10 47 00.8 +12 57 34 & 18.6 & ~~843 (1)~~~~ & ~23 & ~~~~0.48 & 2.3 & 14.0 & 11.1 & 7.14~~~~ & 2.25~~~~~~~~\\
201959 & FS 23 & 10 47 27.5 +13 53 22 & 17.79 & 3009 (3)~~~~ & ~61 & ~~~~0.32 [0.26] & 1.8 & 7.2 & 45.7 & 8.21~~~~ & 1.76~~~~~~~~\\
200592\tablenotemark{d} & P032327 & 10 48 43.3 +12 18 55 & 17.51 & ~~876 (4)~~~~ & ~44 & ~~~~0.26 & 1.1 & 11.9 & 11.1 & 6.88~~~~ &
1.56~~~~~~~~\\
202028 & FS 40 & 10 49 37.0 +11 21 04 & 18.0 & -~~~~~~~~~ & - & ~~~~~~~~- & 2.0 & - & - & $<$ 6.95~~~~ & $<$ 1.81~~~~~~~~\\
201963\tablenotemark{e} & P1424345 & 10 49 52.2 +13 09 42 & 20.0 & ~~766 (2)~~~~ & ~19 & ~~~~1.12 [1.43] & 2.1 & 38.3 & 11.1 & 7.51~~~~ & 3.17~~~~~~~~\\
202029 & LeG23 & 10 50 09.1 +13 29 00 & 19.1 & -~~~~~~~~~ & - & ~~~~~~~~- & 1.7 & - & - & $<$ 6.86~~~~ & $<$ 2.16~~~~~~~~\\
201991 & KK 96 & 10 50 27.0 +12 21 39 & 18.3 & -~~~~~~~~~ & - & ~~~~~~~~- & 1.8 & - & - & $<$ 6.89~~~~ & $<$ 1.88~~~~~~~~\\
202030\tablenotemark{f} & LeG26 & 10 51 21.1 +12 50 57 & 17.2 & -~~~~~~~~~ & - & ~~~~~~~~- & 1.7 & - & - & $<$ 6.86~~~~ & $<$ 1.41~~~~~~~~\\
202031 & LeG27 & 10 52 20.1 +14 42 25 & 18.6 & -~~~~~~~~~ & - & ~~~~~~~~- & 1.7 & - & - & $<$ 6.86~~~~ & $<$ 1.98~~~~~~~~\\
202032 & LeG28 & 10 53 00.7 +10 22 44 & 18.3 & -~~~~~~~~~ & - & ~~~~~~~~- & 1.6 & - & - & $<$ 6.85~~~~ & $<$ 1.84~~~~~~~~\\
202033 & D640-16 & 10 55 03.6 +14 05 35 & 18.6 & 2094 (2)~~~~ & ~40 & ~~~~0.22 & 1.0 & 10.9 & 32.5 & 7.74~~~~ & 1.93~~~~~~~~\\
202034 & D640-12 & 10 55 55.3 +12 20 22 & 18.4 & ~~847 (2)~~~~ & ~22 & ~~~~0.10 & 0.8 & 8.6 & 11.1 & 6.46~~~~ & 1.51~~~~~~~~\\
202035\tablenotemark{g} & D640-13 & 10 56 13.9 +12 00 37 & 17.66 & ~~989 (2)~~~~ & ~28 & ~~~~1.55 [1.69] & 2.0 & 44.7 & 11.1 &
7.65~~~~ & 2.40~~~~~~~~\\
202036 & D640-14 & 10 58 10.5 +11 59 56 & 18.5 & -~~~~~~~~~ & - & ~~~~~~~~- & 0.9 & - & - & $<$ 6.58~~~~ & $<$ 1.67~~~~~~~~\\
202037 & LeG32 & 10 59 17.4 +15 05 07 & 18.7 & 2105 (2)~~~~ & ~42 & ~~~~0.67 & 1.7 & 19.0 & 32.9 & 8.23~~~~ & 2.44~~~~~~~~\\
202038 & LeG33 & 11 00 45.2 +14 10 19 & 18.6 & -~~~~~~~~~ & - & ~~~~~~~~- & 1.7 & - & - & $<$ 6.87~~~~ & $<$ 1.99~~~~~~~~\\
202039\tablenotemark{h} & D640-08 & 11 00 51.9 +13 52 51 & 17.0 & -~~~~~~~~~ & - & ~~~~~~~~- & 0.8 & - & - & $<$ 6.52~~~~ & $<$ 1.01~~~~~~~~\\
202040 & LeG35 & 11 03 02.1 +08 02 53 & 18.1 & 1358 (1)~~~~ & 103 & ~~~~2.01 [1.77] & 2.4 & 25.5 & 17.5 & 8.16~~~~ & 2.65~~~~~~~~\\
\hline
\enddata
\tablenotetext{a}{possible Ring detections; see Figure \ref{ringcont}}
\tablenotetext{b}{SDSS gives cz=1013 km/s which matches HI cz}
\tablenotetext{c}{NED gives cz=617 km/s which matches HI cz}
\tablenotetext{d}{Ring detection, SDSS gives cz=16,775 km/s for optical galaxy}
\tablenotetext{e}{Ring detection, also in ADBS at 754 km/s, SDSS gives cz=53,213.1 km/s for optical galaxy}
\tablenotetext{f}{SDSS gives cz=1019.29 km/s}
\tablenotetext{g}{SDSS gives cz=629.56 km/s with very low s/n}
\tablenotetext{h}{SDSS gives cz=1588.90 km/s}
\end{deluxetable}
\clearpage
\end{landscape}

\end{document}